\documentclass{article}

\usepackage{setspace}

\usepackage{arxiv}
\usepackage[font={small,it}]{caption}
\usepackage[utf8]{inputenc} 
\usepackage[T1]{fontenc}    
\usepackage{hyperref}       
\usepackage{url}            
\usepackage{booktabs}       
\usepackage{amsfonts}       
\usepackage{nicefrac}       
\usepackage{microtype}      
\usepackage{lipsum}
\usepackage{graphicx}
\usepackage{amsmath}
\usepackage[dvipsnames]{xcolor}
\graphicspath{ {./} }
\title{Boosting Resolution and Recovering Texture of micro-CT Images with Deep Learning}

\author{
  Ying Da~Wang \\
  School of Minerals and Energy Resources Engineering\\
  University of New South Wales\\
  Sydney, NSW, 2052 \\
  \texttt{yingda.wang@unsw.edu.au} \\
   \And
  Ryan T.~Armstrong \\
  School of Minerals and Energy Resources Engineering\\
  University of New South Wales\\
  Sydney, NSW, 2052 \\
  \texttt{ryan.armstrong@unsw.edu.au} \\
  \And
  Peyman~Mostaghimi \\
  School of Minerals and Energy Resources Engineering\\
  University of New South Wales\\
  Sydney, NSW, 2052 \\
  \texttt{peyman@unsw.edu.au} \\
}

\begin{document}
\maketitle

\begin{abstract}
Digital Rock Imaging is constrained by detector hardware, and a trade-off between the image field of view (FOV) and the image resolution must be made. This can be compensated for with super resolution (SR) techniques that take a wide FOV, low resolution (LR) image, and super resolve a high resolution (HR), high FOV image. A Super Resolution Convolutional Neural Network (SRCNN) that excels in capturing edge details is coupled with a Generative Adversarial Network (GAN) to recover high frequency texture details. The Enhanced Deep Super Resolution Generative Adversarial Network (EDSRGAN) is trained on the Deep Learning Digital Rock Super Resolution Dataset (DeepRock-SR), a diverse compilation of raw and processed $\mu$CT (micro-Computed Tomography) images. The dataset consists of 12000 sandstone, carbonate, and coal samples of image size 500x500 as a representation of fractured and granular media. The SRCNN network shows comparable performance of  +3-5dB in pixel accuracy (50\% to 70\% reduction in relative error) over bicubic interpolation. GAN performance in recovering texture shows superior visual similarity compared to normal SRCNN and other interpolation methods. Difference maps indicate that the SRCNN section of the SRGAN network recovers large scale edge (grain boundaries) features while the GAN network regenerates perceptually indistinguishable high frequency texture. Extrapolation of the learned network with external samples remains accurate and network performance is generalised with augmentation, showing high adaptability to noise and blur. HR images are fed into the network, generating HR-SR images to extrapolate network performance to sub-resolution features present in the HR images themselves. Results show that under-resolution features such as dissolved minerals and thin fractures are regenerated despite the network operating outside of trained specifications. Comparison with Scanning Electron Microscope (SEM) images shows details are consistent with the underlying geometry of the sample. Recovery of textures benefits the characterisation of digital rocks with a high proportion of under-resolution micro-porous features, such as carbonate and coal samples. Images that are normally constrained by the mineralogy of the rock (coal), by fast transient imaging (waterflooding), or by the energy of the source (microporosity), can be super resolved accurately for further analysis downstream. The neural network architecture and training methodology presented in this study (and SRGANs in general) offer the potential to generate HR $\mu$CT images that exceed the typical imaging limits. 
\end{abstract}


\section{Plain Language Summary}
When capturing an image of the insides of a rock sample (or any opaque object), hardware limitations on the image quality and size exist. This limitation can be overcome with the use of machine learning algorithms that "super-resolve" a lower resolution image. Once trained, the machine algorithm can sharpen otherwise blurry features, and regenerate the underlying texture of the imaged object. We train such an algorithm on a large and wide array of digital rock images, and test its flexibility on some images that it had never seen before, as well as on some very high quality images that it was not trained to super-resolve. The results of training and testing the algorithm shows a promising degree of accuracy and flexibility in handling a wide array of images of different quality, and allows for higher quality images to be generated for use in other image-based analysis techniques.

\section{Introduction}
X-ray micro-computed tomography ($\mu$CT) techniques can generate 3D images that detail the  internal micro-structure of porous rock samples. It allows samples to be used for other analyses, as it does not physically interact with the sample \cite{lindquist, Hazlett, Wildenschild, RN7}. It assists in the determination of the petrophysical and flow properties of rocks \cite{DGDD, pfvs, wang2019multi, peymanK, Krakowska, ferrand1992effect}, as they can resolve features down to a few micrometres over a field of view (FOV) spanning a few millimetres, sufficient enough to characterise the micro-structure of conventional rocks \cite{Flannery, Coenen}. For use in flow simulation \cite{jiang2013representation, jiang2013representation, schluter2014image} and other numerical methods \cite{yi2017pore}, an imaged rock sample must be of sufficiently high resolution to resolve the pore space features and connectivity while also spanning a wide enough FOV to represent bulk properties of the rocks in a way that captures the heterogeneity of the sample \cite{Li_and_Teng}. This is challenging for certain cases of more unconventional rocks such as carbonate rocks that can have up to 50\% of the pore space unresolvable under the resolution of the imaging device \cite{BULTREYS201536} or coal samples that are limited to 15-25 $\mu$m resolutions due to sample fragility resulting in physical sample size constraints causing under-resolution of fracture features \cite{coalFragility}. The capabilities of the scanning device and the physical characteristics of rocks themselves are limiting factors in capturing high resolution $\mu$CT digital rock images, with trade-offs between resolution and field of view. These problems can be addressed by applying super resolution methods on low resolution data, resulting in a high resolution image with a wide FOV that  captures more of the pore space geometry in higher detail \cite{wang2019super}. 

Super resolution methods aim to generate a super resolution (SR) image from only a single low resolution (LR) image (as opposed to using multiple LR images) in a way that most closely resembles its true high resolution (HR) counterpart \cite{sr2003overview}. It is an ill-posed, indeterminate inverse problem with an infinite solution space \cite{SRCNNDong}. Simple, typical methods apply interpolation (bicubic, linear, etc.) on the LR data to generate an higher resolution image that struggles to recover features and is mostly a blurry image with more pixels. More complex methods apply a reconstruction based method that utilises prior knowledge of the domain characteristics to reduce the size of the solution space while generating sharp details \cite{softCuts, gradProfPrior, gradProfSharp}. The scale factor is limited in these methods and the computational time required during image generation is high. Example based methods, also known as learning based methods, utilise machine learning algorithms to improve the speed and performance of SR methods. The statistical relationship between LR and HR features are deduced from a large example dataset, and have been applied with the Markov Random Field \cite{MRFExmaple}, Neighbour Embedding \cite{neighbourEmbedding}, sparse coding methods \cite{sparseRep}, and Random Forest methods \cite{randomForest}. These learning based methods have also been combined with reconstruction based methods to further improve SR performance \cite{unifiedSR}. The sparse coding example based SR technique, while inflexible, has been shown to outperform bicubic interpolation and has since been generalised \cite{SRCNNDong} into the structurally analogous but highly flexible Super Resolution Convolutional Neural Network (SRCNN) for super resolving individual photographic images \cite{SRCNNDong, WangEEDS, EDSR, WDSR, VDSR, SRGANledig}, medical images \cite{umeharaSRCNNmedical, circleGAN, geWangRecon}, and digital rock images \cite{wang2019super, wang3DSRCNN}. Results from SRCNN have shown to consistently outperform previously utilised learning based and reconstruction based methods. Specifically for digital rock images, SRCNN methods have been applied to $\mu$CT images of rocks using a variety of different architectures in 2D and 3D, producing compelling results that can be applied to Digital Rock workflows. 

Using Convolutional Neural Networks to super resolve an image was originally done with a network that applied 3-5 activated convolutional layers \cite{SRCNNDong} to a bicubically upscaled LR image to recover the HR details on a direct mapping. This initial architecture coined the acronym "SRCNN", and was iteratively improved upon with deeper, more complex networks with improved LR to HR mappings. Integrated LR to HR SRCNNs that do not require bicubic upsampling as a processing step are favoured as the need for a fully sized input image increases the computational cost. Deconvolutional layers at the end of the network \cite{FSRCNN} have been replaced by subpixel convolution to reduce checkerboard artifacting \cite{subpixelConv, deconvCheckerboard}. Deeper networks have shown improved results by using the skip connection that adds outputs from shallow layers to deeper layers to preserve important shallow feature sets and improve gradient scaling \cite{VDSR}. Batch normalisation is featured in many deep networks \cite{SRGANledig}, but has since been empirically observed to reduce the accuracy of SRCNN methods that rely on mini batches of cropped images due to the highly variant batch characteristics. As such they were removed in later networks \cite{EDSR}, and are no longer present in more recent formulations \cite{WDSR}. The typical SRCNN algorithm attempts to generate a SR image that has the smallest L2 (Mean Squared Error) or L1 \footnote{Lebesgue spaces, named after Henri Lebesgue, L1 and L2 are specific cases of $L^p$ spaces} (Mean Absolute Error) loss across the image \cite{cycleGAN} which has been found to be more robust against outlying features and improve convergence rates. The generated SR images tend to have high PSNR values but a low perceptual quality, with high frequency texture features lost as smearing occurs in order to achieve a high "average" accuracy that represents multiple possible HR realisations \cite{perceptLoss}, a manifestation of the local minima problem in neural network training. While larger scale features that have high contrast edges over multiple pixels are recovered by SRCNN, the overall image fidelity is unsatisfying due to the loss of texture. Efforts to address this problem have resulted in the use of hybrid loss functions that combine a pixelwise L1 or L2 loss with a featurewise loss that is calculated as the L2 loss of features extracted from a convolutional layer in a pretrained model \cite{perceptLoss2}. This has shown superior perceptual results compared to standard loss functions. This problem of texture loss has been most successfully addressed by the introduction of the SRGAN network.

Generative Adversarial Networks (GANs) \cite{ganGoodfellow} are comprised of a pair of neural networks, a generator and a discriminator, being participants in a zero-sum game. The discriminator learns to classify inputs as either real or fake, while the generator attempts to generate fake inputs to the discriminator that are good enough for it to classify as real. In its simplest form, random noise is fed into a generator that is trained to transform the input noise into data that closely resembles the real data from the training set. The discriminator is trained alongside the generator to distinguish between the real training set data, and the generated data. As the neural networks are trained, the discriminator becomes better at identifying fake data, while the generator becomes better at generating data that fools the discriminator. By combining a SRCNN network (the generator) with an image classification network (the discriminator), an SRGAN is formed. 

SRGANs have been applied to the generation of photo-realistic and highly textured images that score highly on human surveys of image quality \cite{SRGANledig, perceptLoss}. Cycle Consistent GANs (cycleGANs) \cite{cycleGAN} have been used as a semi-supervised method of generating SR images, by training on unpaired SR and LR datasets \cite{circleGAN}. While SRGAN results in features that look to the human eye as realistic when surveyed, the resulting generated SR images are lower in pixelwise accuracy compared to SRCNN due to pixelwise mismatch \cite{circleGAN}. Since $\mu$CT images contain significant amounts of image noise and texture as high frequency features, this is also recovered inadvertently recovered. As image segmentation is common in most Digital Rock workflows \cite{iassonov2009segmentation}, SRCNN tends to suffice as they possess some form of intrinsic noise suppression while maximising edge recovery \cite{wang2019super}. While sufficient for conventional samples that are resolvable at the 3-5 $\mu$m resolution scale, the blurring of intraphase texture is detrimental to images such as carbonate and coal, with significant under-resolution features. In the context of digital rock imaging, to the authors' best knowledge, SRGAN networks have not yet been applied to the generation of SR $\mu$CT images. 
 
In the following sections of this paper, we introduce the DeepRock-SR dataset used to train and validate the results from the Super Resolution Generative Adversarial Network, which in this case, the generator (acting as the SRCNN) is  a modified EDSR \cite{EDSR} network (seen in Figure \ref{fig:edsr}), and the discriminator is a deep convolutional classifier (seen in Figure \ref{fig:discrimNet}). The generator shows comparative performance to other SRCNN networks, with performance on the DeepRock-SR dataset is in line with expected improvements in PSNR of +3-5 dB, which translates to a reduction in relative pixelwise error of 50\% to 70\% respectively according to Eqn \ref{eqn:PSNR}. Difference maps of the validation sample images confirm SRCNN recovery of bulk features, while the SRGAN network is able to regenerate the high resolution texture. Extrapolation of network performance with unseen external images incurs minimal performance loss, while application of image augmentation to generalise the model capabilities indicates good adaptation to image noise and blur. Super resolution images of the high resolution sandstone, coal, and carbonate samples in the DeepRock-SR dataset are also generated. This extrapolates the network performance past the image resolutions of the training (3-5 $\mu$m) to under 1 $\mu$m to observe the improvement in sub-pixel resolution past the physical capabilities of $\mu$CT scanners, showing excellent visual regeneration of under-resolution features across a wide range of image resolutions, which are also confirmed with a direct comparison with high fidelity SEM images.

\pagebreak

\section{Materials and Methods}
\label{sec:Materials and Methods}

\subsection{Datasets and Digital Rocks}
\label{sec:Datasets and Digital Rocks}

\begin{figure}[htp!]
  \centering
    \includegraphics[width=\textwidth]{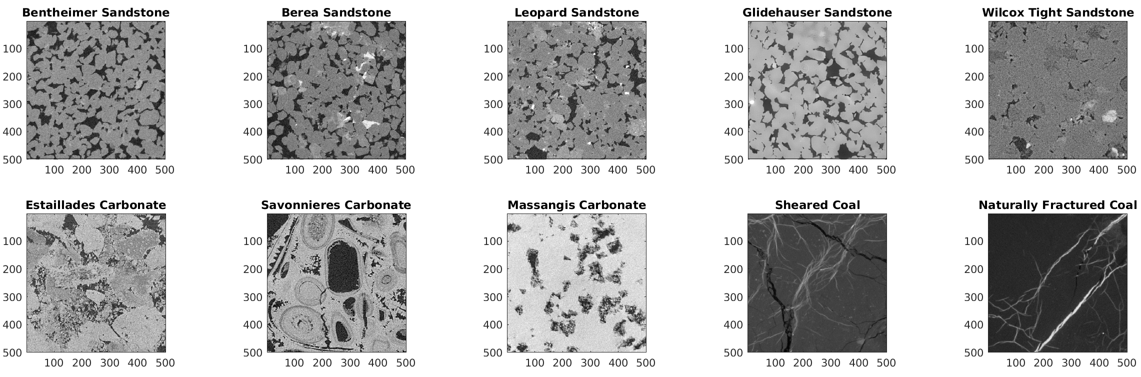}
    \caption{Sample slices of images from DeepRock-SR}
    \label{fig:sampleImages}
\end{figure}

The SRGAN network in this study is trained on the DeepRock-SR-2D dataset \cite{DRSRD3}, which comprises of 12000 500x500 high resolution unsegmented slices of various digital rocks of sandstone, carbonate, and coal, with image resolution ranging from 2.7 to 25 $\mu$m, outlined in Table \ref{tab:DRSRD3Table}. Each rock type is represented by 4000 images for an even distribution of rock geometries. The images are shuffled and split 8:1:1 into training, validation, and testing sets. From a pore morphology and image characteristics perspective, the rock types differ greatly from each other, seen in Figure \ref{fig:sampleImages}. The samples in the dataset are diverse, and represent both simple resolved sandstone images (sandstone), and complex under-resolved images of carbonate microporosity and coal fracture networks. There is also a Gildehauser sandstone sample that is highly processed by Non-Local Means Filtering \cite{nonLocal}, resulting in a very smooth intraphase image. The carbonate images are under-resolved due to much of the pore space measuring under 1 $\mu$m in resolution \cite{BULTREYS201536} (requiring an image resolution approaching 100 nm to adequately resolve such features). The coal samples are inadequately resolved due to the fragility of the rock itself, restricting the size to a larger than usual sample, limiting the resolution of the resulting image \cite{coalFragility}. The images are downsampled by a factor of 2x and 4x using the matlab imresize function, used commonly for SR dataset benchmarking \cite{wang2019super}, with one set being downsampled exclusively by bicubic interpolation, while another set is downsampled with random kernels of either box, triangle, lanczos2, and lanczos3. In this paper, to generalise the results, the set with random (unknown) downsampling functions is used for training at a scale factor of x4. All training and testing was performed on a GTX1080ti Nvidia GPU with TensorFlow 1.12. While the training and testing dataset used here is of a fixed, relatively small size per image, the resulting trained SRCNN and SRGAN networks are "fully convolutional networks" that can be applied to any image of any size, without cropping or windowing.

\begin{table}[htp!]
 \caption{DeepRock-SR Dataset constituents. A diverse set of digital rock images.}
 \label{tab:DRSRD3Table}
  \centering
  \begin{tabular}{lllll}
    \toprule
    \multicolumn{2}{c}{Part}                   \\
    \cmidrule(r){1-2} 
    Name & Source & Resolution ($\mu$m) & Processing & Reference \\
    \midrule
    Bentheimer Sandstone 1 & UNSW Tyree X-ray & 3.8 & None & \cite{DRSRD1} \\
    Bentheimer Sandstone 2 & ANU CTLab & 4.9 & None & \cite{herringSands}\\
    Berea Sandstone & ANU CTLab & 4.6 & None & \cite{herringSands}\\
    Leopard Sandstone & ANU CTLab & 3.5 & None & \cite{herringSands}\\
    Gildehauser Sandstone & TOMCAT beamline & 4.4 & Non-Local Means Filtered & \cite{glideSand}\\
    Wilcox Tight Sandstone & UT Austin & 2.7 & None & \cite{wilcox}\\
    \midrule
    Estaillades Carbonate & UGCT HECTOR & 3.1 & None & \cite{estCarb}\\
    Savonnieres Carbonate & UGCT HECTOR & 3.8 & None & \cite{savCarb}\\
    Massangis Carbonate & UGCT HECTOR & 4.5 & None & \cite{massCarb}\\
    \midrule
    Sheared Coal &  Ultratom RXSolutions & 25 & None & \cite{shearedCoal}\\
    Naturally  Fractured Coal & Ultratom RXSolutions & 25 & None & \cite{fracCoal}\\
    \bottomrule
  \end{tabular}
\end{table}

While datasets are generated synthetically to remain consistent with other SR datasets, real LR $\mu$CT images are not mapped as consistently to HR equivalents, and registration error of even a few pixels can considerably reduce training results if the network requires images to be perfectly matched. Pixel binning of greyscale values \cite{Guan2019} has been used to model LR digital rock images, but in a practical case, with a lower energy detector or a shorter exposure time, the resulting LR $\mu$CT image can have different noise and blur compared to a synthetic LR images downsampled from a HR registered equivalent. Aside from the images contained within the DeepRock-SR dataset used for training and validation, external samples are also used in this study to extrapolate network performance. These are outlined in Table \ref{tab:externalSamples}. 

\begin{table}[htp!]
 \caption{External digital rock images used to extrapolate network performance}
 \label{tab:externalSamples}
  \centering
  \begin{tabular}{lllll}
    \toprule
    \multicolumn{2}{c}{Part}                   \\
    \cmidrule(r){1-2} 
    Name & Source & Resolution ($\mu$m) & Processing & Reference \\
    \midrule
    Bentheimer Sandstone & Equinor & 7.0 & None & \cite{bentRam} \\
    \midrule
    Ketton Carbonate & Imperial College & 5.0 & None & \cite{kettonCarb}\\
    \bottomrule
  \end{tabular}
\end{table}

\subsection{Super Resolution Convolutional Neural Network}
\label{sec:SRCNN}

SRCNN networks generate an SR image from only an LR image through a feed-forward CNN, the generator $G$. Here, $G$ is composed of multiple layers $L$ of weights $w_L$ and biases $b_L$ such that $SR=G_{w_L, b_L}(LR)$, that are calculated by optimising an objective loss function $f_{Loss}(SR, HR)$. The LR image input can be described as a tensor of shape $(N_x, N_y, N_c)$, where $N_x$ and $N_y$ are the dimensions of the image, and $N_c$ are the number of colour channels present. At a scale factor of 4 used in this study, the resulting SR image dimensions are $(4N_x, 4N_y, N_c)$. 

\begin{figure}[htp!]
  \centering
    \includegraphics[width=\textwidth]{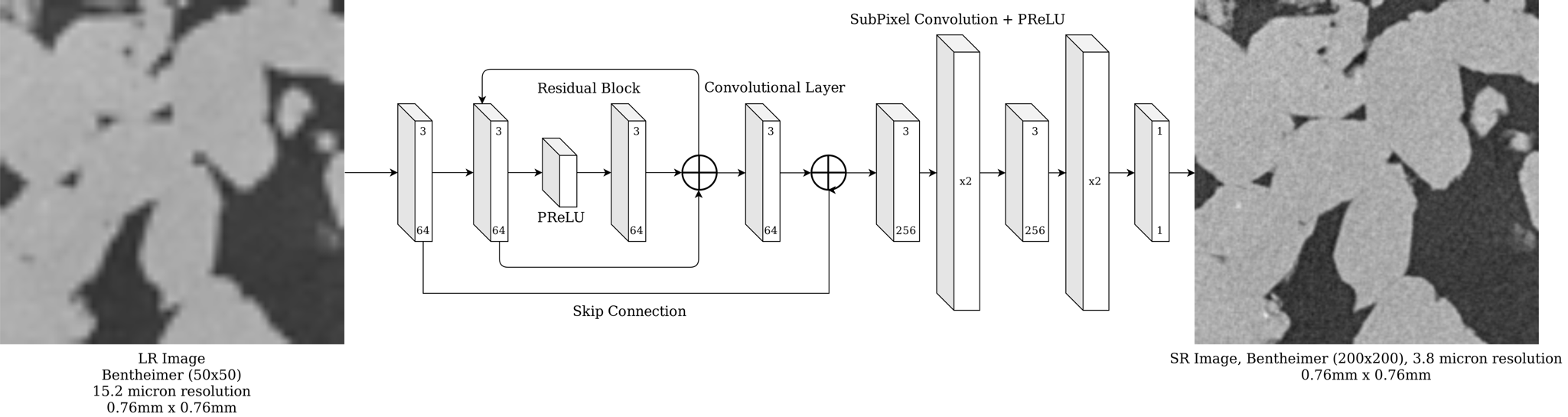}
    \caption{Architecture of the modified EDSR network. All ReLU activation layers are replaced with PReLU layers, and the subpixel convolutions are activated by PReLU for additional non-linearity during the critical upscaling layers}
    \label{fig:edsr}
\end{figure}

The architecture used in this study is based on the Enhanced Deep Super Resolution (EDSR) \cite{EDSR} and the SR-Resnet \cite{SRGANledig} networks. In particular, the overall EDSR architecture is retained, with the Rectified Linear Unit (ReLU) layers replaced by Parametric Rectified Linear Unit (PReLU) layers from SR-Resnet. These steps are taken as the use of batch normalisation layers in SRCNN networks have been shown to be detrimental to training \cite{WDSR}, while the PReLU layers provide a minor improvement in performance for minimal computational impact \cite{prelu}. In this paper, the EDSR network utilises convolutional layers of kernel size 3 with 64 filters over 16 residual layers and skip connections throughout the layers prior to upscaling, as seen in Figure \ref{fig:edsr}. Typically, EDSR and similar SRCNN networks use the L2 loss as the objective minimisation function, defined as the pixelwise mean of the sum of the squares of the differences between the generated SR and ground truth HR images:

\begin{equation}
\label{eqn:L2}
    L_{2_{Loss}}=\frac{\sum_i(SR_i-HR_i)^2}{\sum_ii}
\end{equation}

This choice of loss function in SRCNN has been superseded by the L1 loss defined as:

\begin{equation}
\label{eqn:L1}
    L_{1_{Loss}}=\frac{\sum_i|SR_i-HR_i|}{\sum_ii}
\end{equation}

as it has been shown to converge SRCNN networks faster, with better results \cite{cycleGAN}. This is suspected to be due to a better representation of the high frequency noise and texture of an image, resulting in a lower local minimum during training. 

The modified EDSR is trained for 100 epochs at 1000 iterations per epoch with mini-batches of 16 cropped 192x192 images using an L1 loss as in Eqn \ref{eqn:L1} with a learning rate of 1e-4 using the Adam optimiser \cite{AdamKingma}. The PSNR metric is tracked during the training and validation of the DeepRock-SR dataset, and is calculated assuming that the HR and SR image pixel values lie between [-1,1] such that $I=2$:

\begin{equation}
\label{eqn:PSNR}
    PSNR=10log_{10}\frac{I^2}{L_{2_{Loss}}}
\end{equation}

Further training may result in a higher final SRCNN-only PSNR score, however previous SRCNN tests on the DRSRD1 dataset \cite{DRSRD1} has indicated that 100 epochs of 1000 iterations of 16x192x192 batches are typically enough to saturate training for digital rocks, with the PSNR plateauing at around 50 epochs using a learning rate of 1e-4. After the training of this SRCNN-only model is done, the SRGAN is initialised and trained.

\subsection{Super Resolution Generative Adversarial Network}
\label{sec:SRGAN}

The GAN network attaches onto the generative SRCNN network in the form of an image classification type network, a discriminator that identifies and labels input images as real and fake. The discriminator is trained on the SR and HR images, gradually becoming better at identifying SR and HR images. As the discriminator improves, its classification feedback is passed back to the generator to allow it to generate progressively more realistic SR images that attempt to fool a progressively improving classifier. 

\begin{figure}[htp!]
  \centering
    \includegraphics[width=\textwidth]{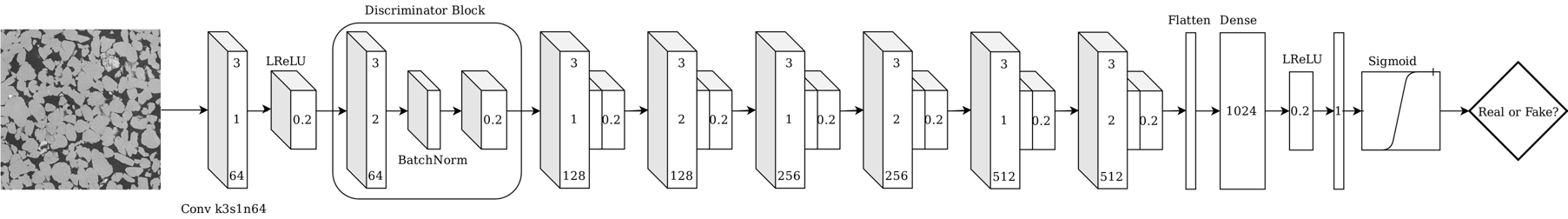}
    \caption{Architecture of the Discriminator network}
    \label{fig:discrimNet}
\end{figure}

The discriminator network is built with 8 discriminator blocks, with each block containing a convolutional layer of kernel size 3, stride 2 and filter numbers increasing from 64 to 512, LReLU activation ($\alpha=0.2$), and batch normalisation as can be seen in Figure \ref{fig:discrimNet}. The network ends with 2 dense layers followed by sigmoid activation to obtain probabilistic values. 

The discriminator ultimately outputs a single value between [0,1] for each image it is classifying, that acts as a logit, or a "probability", that the image is either real (1) or fake (0). These probabilities are discretised appropriately when classifying images. The SR and HR images are labelled as $y_{HR}=1$ and $y_{SR}=0$, while the binary cross entropy (BXE) loss function of the discriminator output ${output}_{discrim} = p(SR,HR)$ is given by:

\begin{equation}
\label{eqn:BXE}
\begin{aligned}
{BXE}_{HR}=-\frac{1}{N}\sum_{i=1}^Ny_{HR}log(p(HR))\\
{BXE}_{SR}=-\frac{1}{N}\sum_{i=1}^Ny_{SR}log(p(SR))
\end{aligned}
\end{equation}

Since the output of the discriminator has passed through a sigmoid function, the Binary Cross Entropy calculated here is also known as the Sigmoid Cross Entropy. The discriminator network is trained with the Adam optimiser and a learning rate of 1e-4 to minimise the total classification loss between real and fake images:

\begin{equation}
\label{eqn:dLoss}
    D_{Loss}={BXE}_{HR}+{BXE}_{SR}
\end{equation}

The GAN begins training after the generator $G$ has reached its prescribed number of iterations (see Section \ref{sec:SRCNN}), and both GAN and SRCNN are trained for 150,000 further iterations. To encourage the generation of texture and features that a typical SRCNN generator would omit, the pixelwise L1 loss is supplemented with a perceptual loss parameter \cite{SRGANledig} calculated as the mean squared error between the SR and HR feature maps obtained as the fully convolutional output from the 16th layer of the 19 layer VGG network \cite{perceptLoss2}. This term, the $VGG_{19_{Loss}}$, corresponds to the 4th convolutional output prior to the 5th max-pooling layer in the VGG-19 network. The feature maps obtained by passing an image through the network, $\phi$, are used to compute the L2 loss as: 

\begin{equation}
\label{eqn:vgg}
    VGG_{19_{Loss}}=\frac{\sum_i(\phi_{SR_i}-\phi_{HR_i})^2}{\sum_ii}
\end{equation}

The generator and discriminator are coupled together by passing the discriminator SR outputs $p(SR)$ into the generator as part of the generative loss function. The adversarial loss term (ADV) takes the form of a BXE against labels of 1:

\begin{equation}
\label{eqn:adv}
    {adv}_{Loss}=-\frac{1}{N}\sum_{i=1}^Nlog(p(SR))
\end{equation}

\begin{figure}[htp!]
  \centering
    \includegraphics[width=\textwidth]{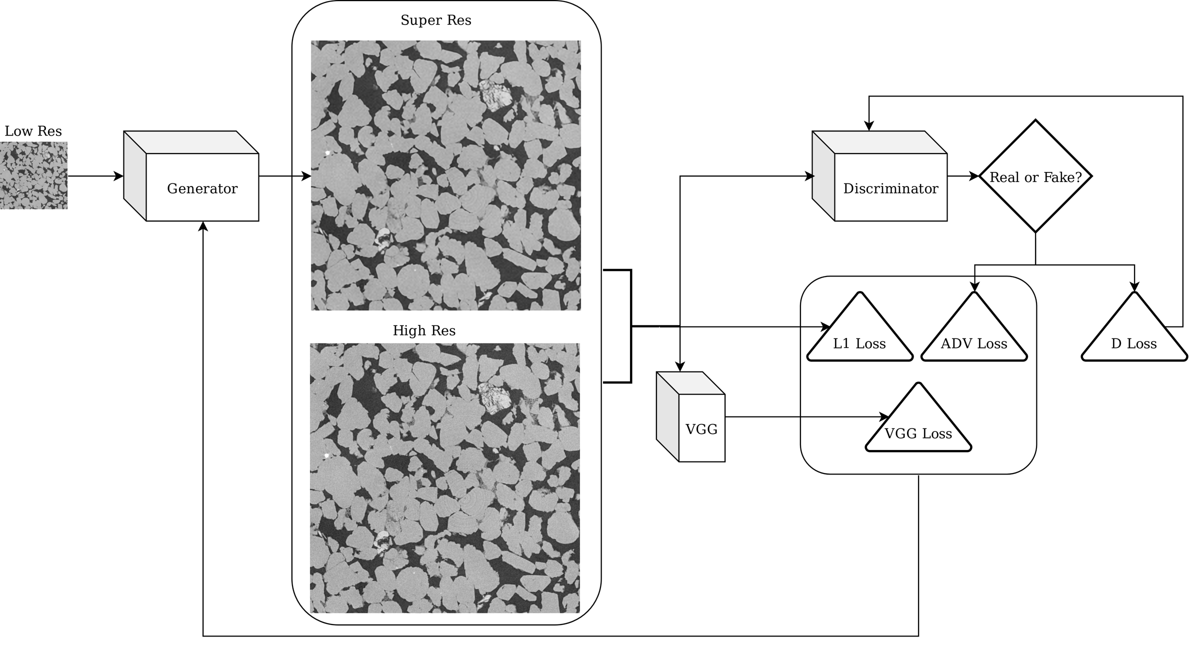}
    \caption{Overall architecture of the network, showing the coupling of Generator and Discriminator Networks}
    \label{fig:SRGANnet}
\end{figure}

This term is minimised when the discriminator classifies the SR images as real images (1 label), despite being trained to classify SR images as fake (0 label) based on Eqn \ref{eqn:BXE}. The generator loss function is now a hybrid function defined as:

\begin{equation}
\label{eqn:gLoss}
    {G}_{Loss}=L_{1_{Loss}}+\alpha VGG_{19_{Loss}}+\beta {adv}_{Loss}
\end{equation}

where $\alpha$ and $\beta$ are scaling terms. In this paper, we take them to equal 1e-5 and 5e-3 respectively to scale near the magnitude of the L1 loss. The choice of the scaling terms will affect the accuracy of the SRGAN network, as the generation of perceptually accurate features by the discriminator works adversarially to the generation of pixelwise accurate features by the L2 loss. As a whole ensemble, the network of generator and discriminator is depicted in Figure \ref{fig:SRGANnet}. 

\pagebreak

\section{Results and Discussion}
\label{sec:Results and Discussion}

\subsection{Training and Validation Metrics}
\label{sec:Training and Validation Metrics}

The network is trained using the first 9,600 images of the shuffled subset of the DeepRock-SR dataset, which is comprised of a selection of $\mu$CT images listed in Table \ref{tab:DRSRD3Table}. The parameters used are outlined in sections \ref{sec:SRCNN} and \ref{sec:SRGAN}. Losses and metrics over the training epochs are tracked over 250 epochs as a moving average with a window of 1000 over 250,000 iterations. The training schedule is comprised of 100 epochs of SRCNN generator training and a further 150 epochs of SRGAN training with the discriminator active.  

\begin{figure}[htp!]
  \centering
    \includegraphics[width=\textwidth]{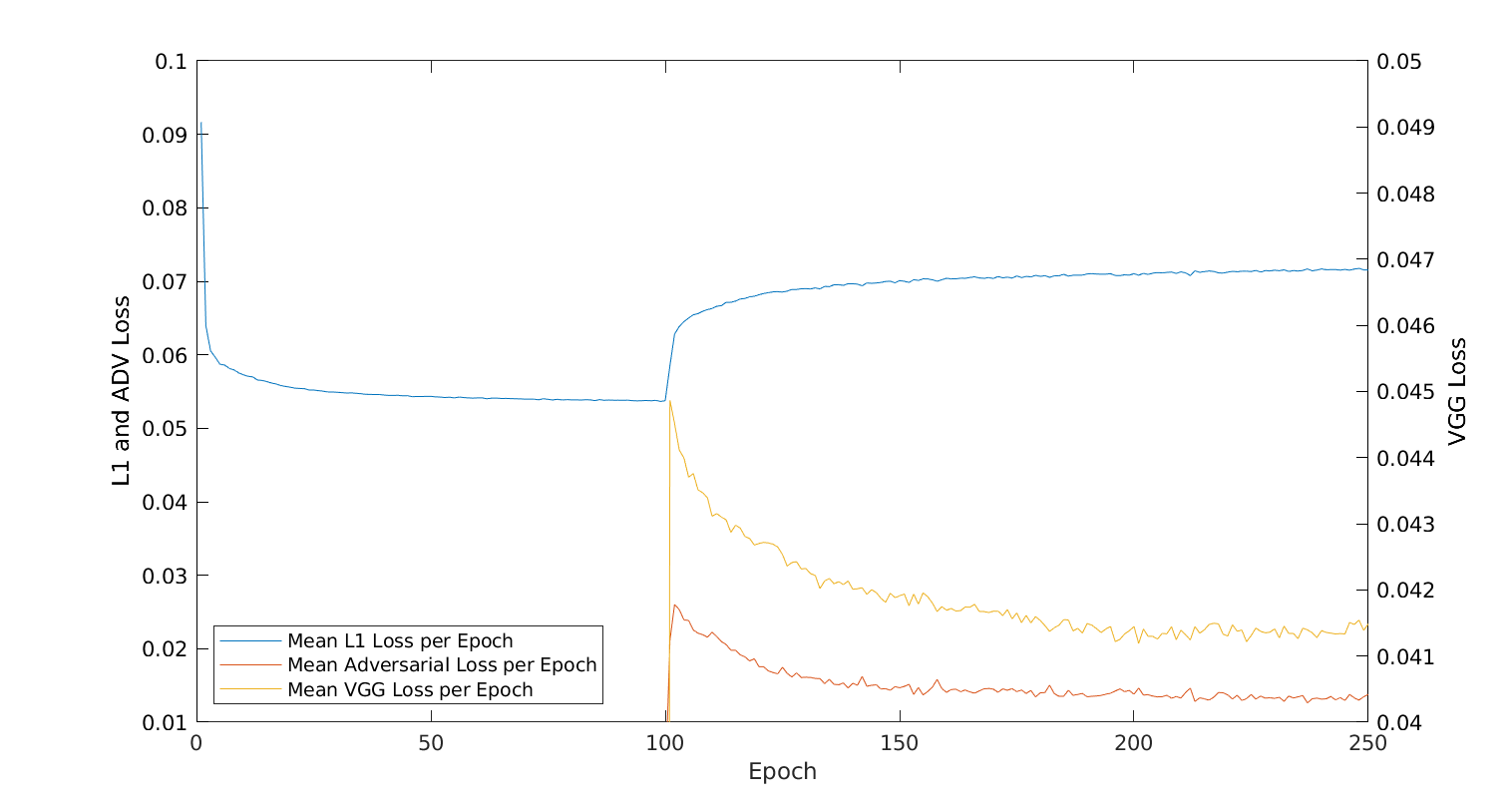}
    \caption{Generative during training of SRGAN. SRCNN is trained with only a pixelwise L1 loss for the first 100 epochs, followed by the activation of SRGAN, with VGG feature losses and adversarial losses falling over epochs 101-250 with an increase in L1 losses.}
    \label{fig:genLosses}
\end{figure}

\begin{figure}[htp!]
  \centering
    \includegraphics[width=\textwidth]{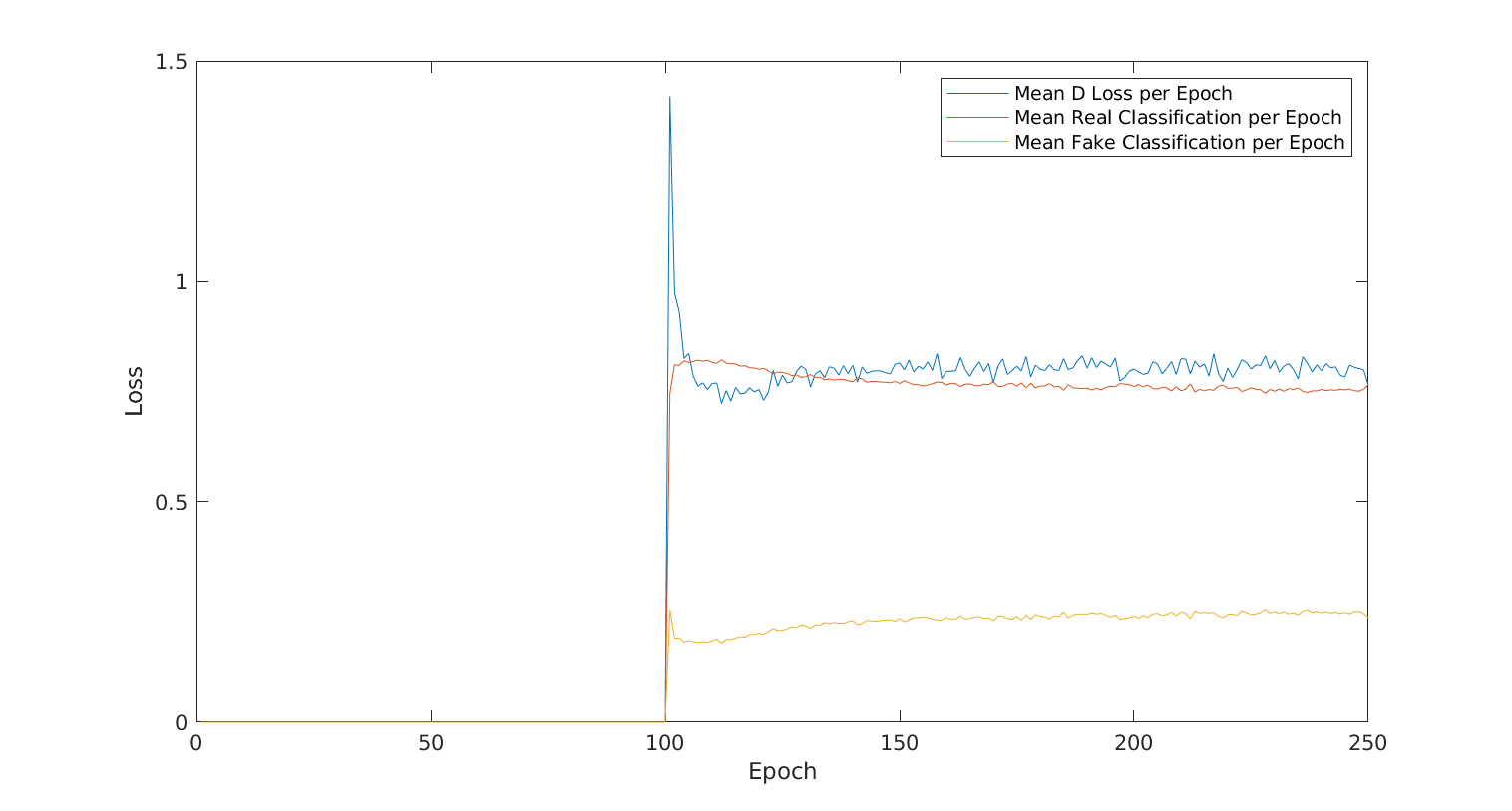}
    \caption{Discriminative losses during training of SRGAN. From epochs 101-250, the discriminator losses rise first fall as it is trained to distinguish SR and HR images, then rises as the generator catches up and begins to fool the discriminator. A perfect generator would result in a $D_{loss}$ of 1, and classification losses of 0.5.}
    \label{fig:discLosses}
\end{figure}

\begin{figure}[htp!]
  \centering
    \includegraphics[width=\textwidth]{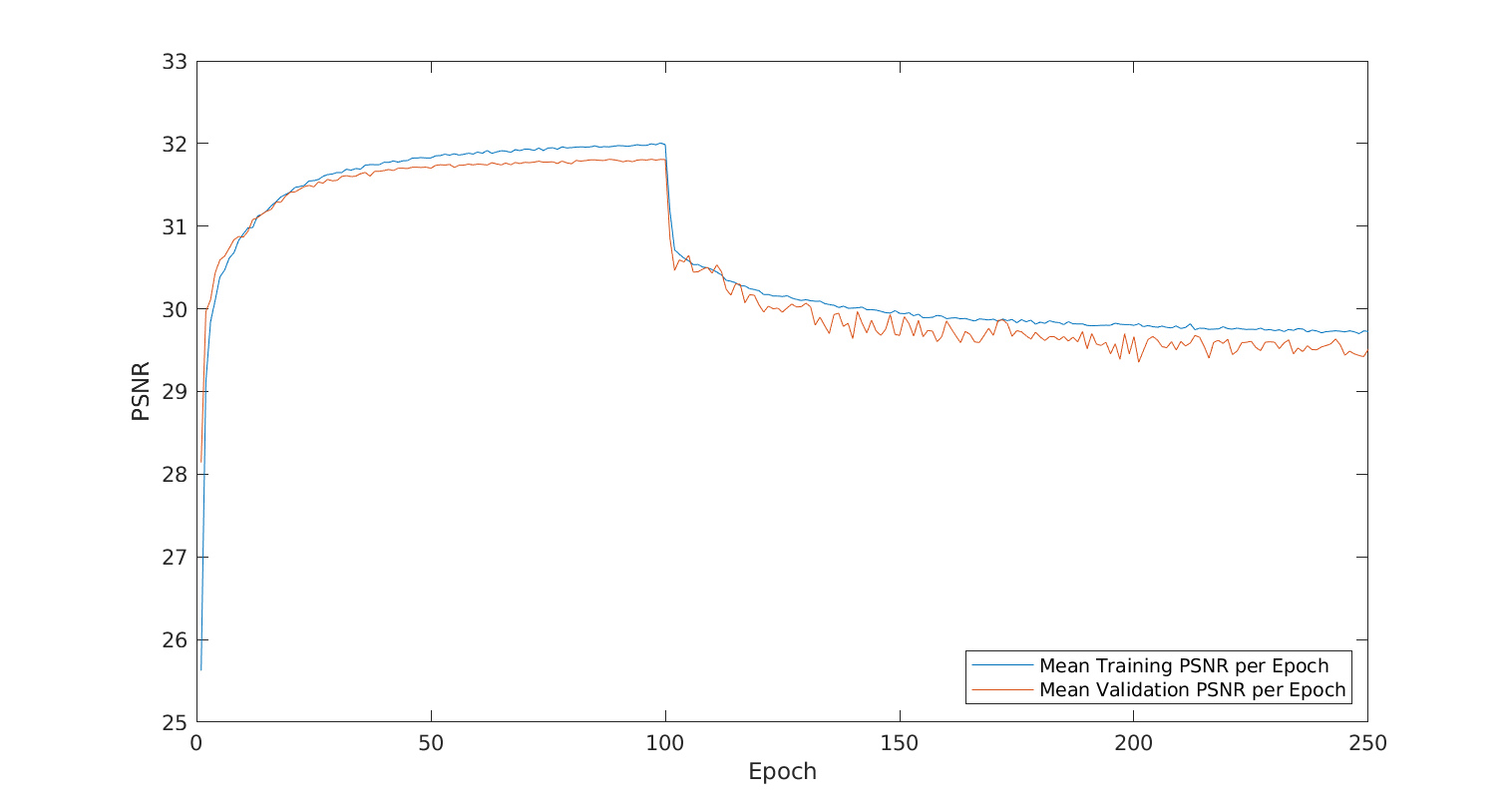}
    \caption{PSNR metrics over the training epochs. Like the L1 loss, the PSNR value falls once the discriminator is activated as texture is regenerated. Training and validation PSNR values track closely, indicating the minibatch window size of 192x192 is sufficient to capture details of the images.}
    \label{fig:trainValPSNR}
\end{figure}

Figure \ref{fig:genLosses} shows the generator losses as described in Eqn \ref{eqn:gLoss}, with the L1 pixelwise loss, the VGG featurewise loss, and the ADV adversarial classification loss. The L1 loss falls to a plateau during the first 100 epochs and results show a 0.053 L1 error between SR and HR images in the training dataset. This translates to a training PSNR of 32, shown in figure \ref{fig:trainValPSNR}, and an equivalent L2 error of .0025 (calculated from Eqn \ref{eqn:PSNR}). These are consistent with typical expected results for SRCNN training on digital rocks \cite{wang2019super}, and will be elaborated upon in the later section \ref{sec:Validation and Testing Analysis}. The VGG and ADV losses are activated after 100 epochs, and fall to a plateau over 150 further epochs. The VGG loss falls from 0.045 to 0.041, while the ADV loss drops from 0.025 to 0.0125. During the SRGAN training, the L1 loss rises to 0.071 with an associated PSNR of 29.7. Similarly during SRGAN training, the discriminator metrics are tracked (see Figure \ref{fig:discLosses}) and show a stable training of the discriminator, plateauing to an overall $D_{loss}$ value of 0.8. The separate classification probabilities $p(HR)$ and $p(LR)$ (see Eqn \ref{eqn:BXE}) are also plotted, and show that the final equilibrium between generator and discriminator results in a classification accuracy of 75\%. 

As the training is run, at the end of every epoch (1000 mini-batch iterations), 1200 fullsize 500x500 images from the DeepRock-SR-2D shuffled dataset are used to validate the generator. The results are plotted with the training PSNR calculated on the 1000 16x192x192 mini-batches in each training epoch, shown in Figure \ref{fig:trainValPSNR}. PSNR values obtained during validation track closely to the training PSNR, with a reduction of 0.2 to 0.3, settling to approximately 29.5 compared to 29.7. The variation in the validation PSNR increases significantly when the SRGAN is active, reflected in the VGG and ADV losses in Figure \ref{fig:genLosses} due to the adversarial interplay between generator and discriminator. From both Figures \ref{fig:trainValPSNR} and \ref{fig:genLosses}, it can be seen that there are potentially further improvements in the SRCNN performance past 100 epochs, as the initial generator training is not completely plateaued when the discriminator is activated.

During SRGAN training, the rise in the L1 loss in order to obtain a lower VGG and ADV is indicative of limitations in the SRGAN network architecture. In an ideal optimisation, the increase in higher frequency texture features should also be reflected in the L1 loss. However, the Adam optimisation \cite{AdamKingma} (or, likely any optimiser) utilised in SRGAN to obtain these texture features are in competition with the L1 loss which tends to favour smoother, averaged pixels. The presence of the VGG loss tends to act as a bridging parameter between the smoothing effect of the L1 loss and the texture generation of the ADV loss \cite{SRGANledig}.

\subsection{Validation and Testing Analysis}
\label{sec:Validation and Testing Analysis}

The validation applied during training of the network (see Figure \ref{fig:trainValPSNR}) is applied to a bulk shuffled collection of 1,200 images from the DeepRock-SR dataset. This is expanded upon in this section to include the 1,200 images in the testing section of the dataset and split apart into sandstone, carbonate, and coal subsets. 

\begin{figure}[htp!]
  \centering
    \includegraphics[width=\textwidth]{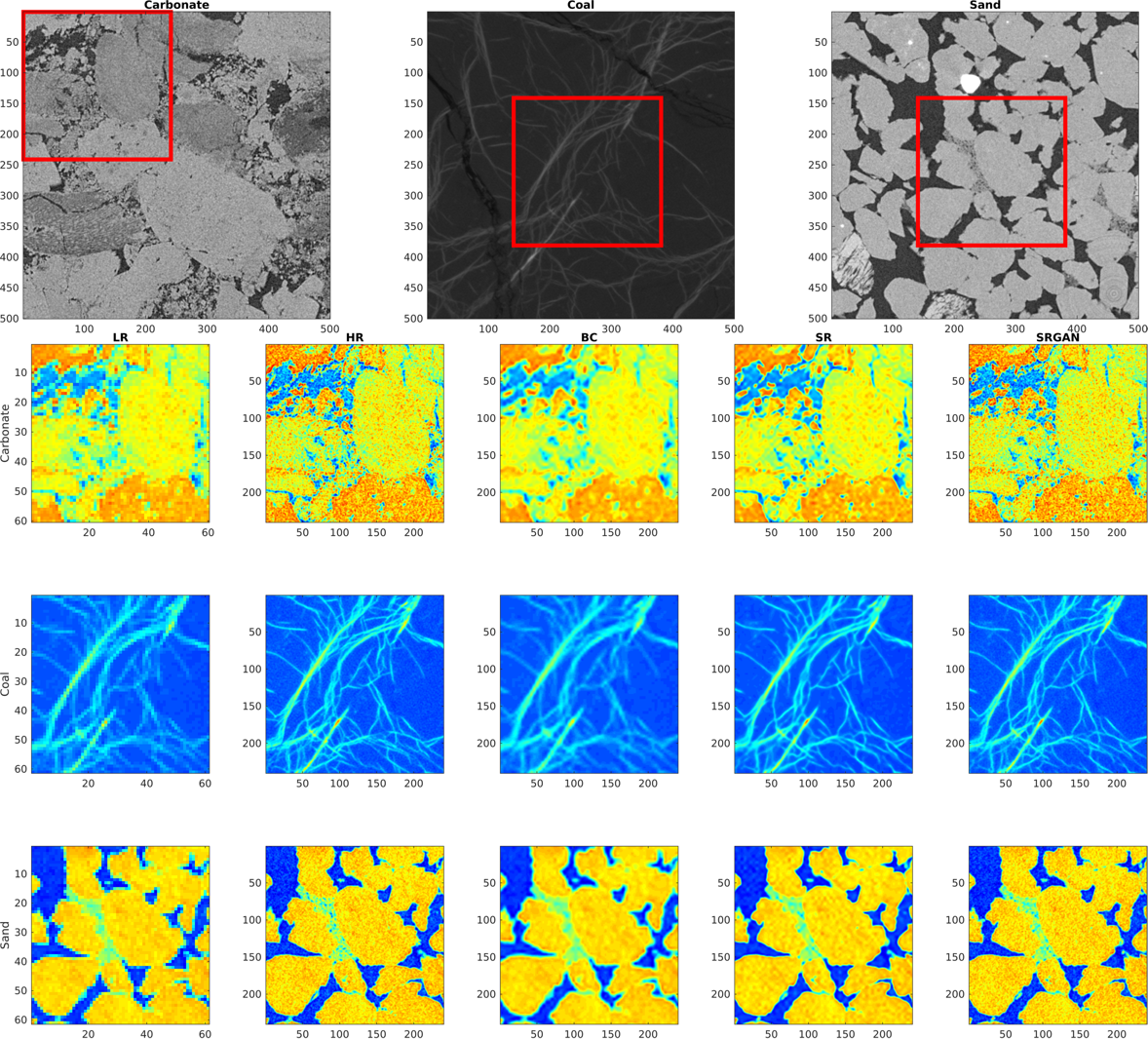}
    \caption{Sample images from the validation of the network, highlighted with a Jet colourmap for visual clarity. From the LR image, a bicubic image is generated, a SR image is generated from the epoch prior to activation of the GAN, and the SRGAN image is generated at the end of the training. Visually, the results show a gradual improvement in feature recovery until the SRGAN images that look perceptually identical to the HR images.}
    \label{fig:trainValImgs}
\end{figure}

\begin{figure}[htp!]
  \centering
    \includegraphics[width=\textwidth]{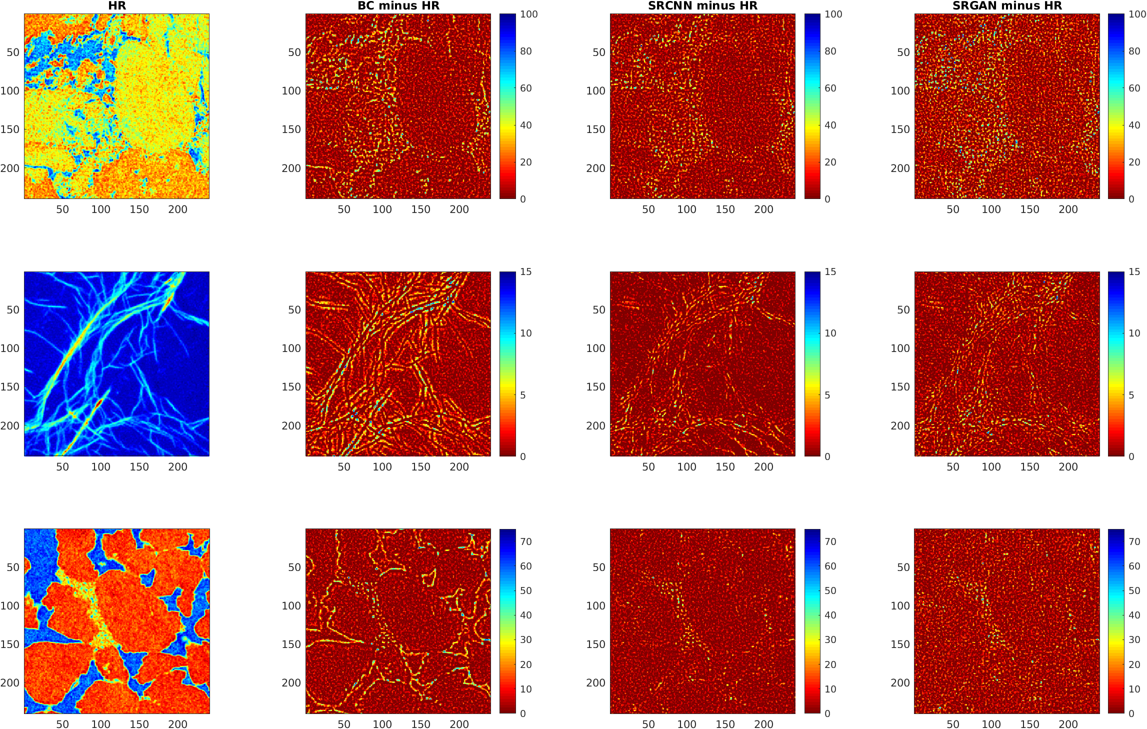}
    \caption{Difference maps of the sample images shown in Figure \ref{fig:trainValImgs}. This more quantitative analysis of the images reveals the benefits and limitations of the SRGAN results. No distinct improvement in pixelwise accuracy in the sandstone and carbonate examples can be seen despite the considerable textural and perceptual improvement. Coal SRGAN images further improve the edge recovery from the SR images likely due to the SRCNN method struggling to cope with the high contrast and thin fracture features.}
    \label{fig:trainValDiffs}
\end{figure}

An initial visualisation of the resulting validation images is shown in Figure \ref{fig:trainValImgs}, and difference maps with respect to the HR images is shown in Figure \ref{fig:trainValDiffs}. The LR image (top row) and HR image (2nd row) are used as comparison points to the bicubically (BC) generated image, SR image, and SRGAN image that are generated only from the LR image. Inspection of these sample validation images indicate that there is an increasing trend of visual sharpness and texture from BC to SR to SRGAN images. While present in all 3 types of rock images, this is especially apparent in the carbonate images that contain highly heterogeneous features such as oolitic vugs and high frequency texture associated with microporosity.  

The difference maps of these sample images, shown in Figure \ref{fig:trainValDiffs}, provide a better indication of the ability for each method to recover important features from a single LR image. The BC images struggle to regenerate large scale features such as the edges of coal fractures and grain contacts in sandstone and carbonates. The SR images lose intragranular detail but recover larger features well, which is useful for cases where the details within each phase are irrelevant. This is typically the case for well resolved grains and pores of sandstones. SRGAN images tend to still struggle to achieve a pixelwise match to the HR image. The perceptual texture and sharpness seen in the images in Figure \ref{fig:trainValImgs} are shown in the difference maps in Figure \ref{fig:trainValDiffs} to not contribute any significant improvement in accuracy. This is most noticeable in the first sand image (from the left), where there exists a region of dissolved microporous mineral that is below the resolution of both the HR and LR images. The SR image results in a mostly blurry and featureless characterisation of this under-resolution area, while the SRGAN recovers very convincingly the mineralised features of this region. The difference maps of this area however, show that the overall pixelwise accuracy has not improved. Aside from the texture regeneration, it can be seen in the coal sample images that SRGAN also improves upon the edge recovery of SRCNN, resulting a better match with the well defined fractured features of the coal images. 

\begin{figure}[htp!]
  \centering
    \includegraphics[width=\textwidth]{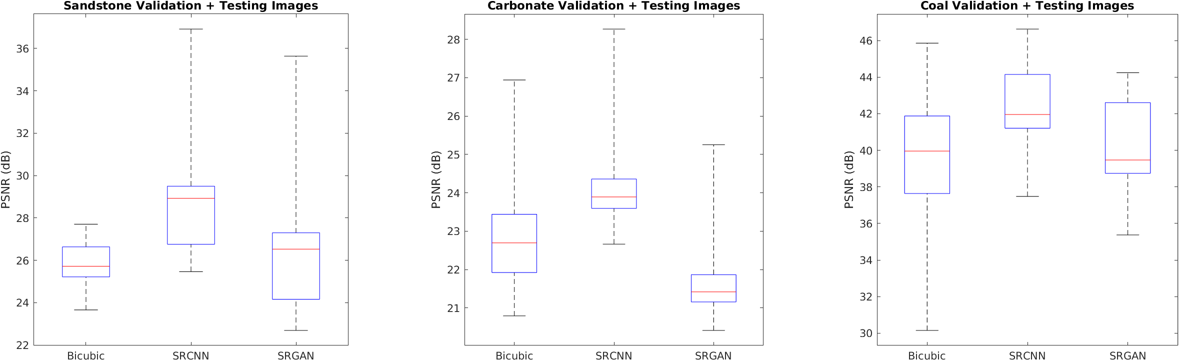}
    \caption{Boxplots of the PSNR values computed by comparison of original HR images to BC, SR, and SRGAN images for sandstone, carbonate, and coal. Results show that SRCNN outperforms bicubic interpolation, while SRGAN performs worse. The texture recovery capabilities of the SRGAN network are not quantifiable by the PSNR metric.}
    \label{fig:boxplotFig}
\end{figure}

Comparative validation of the SRCNN and SRGAN results is performed on separated subsets of sandstone, carbonate, and coal images, each comprising 800 500x500 images. The PSNR metrics for each subset are computed and shown in Figure \ref{fig:boxplotFig}, and show that SR images outperform bicubic interpolation, while SRGAN images are less accurate on a pixelwise basis. The overall PSNR boosts are given in Table \ref{tab:valTestPSNRTable}, and show similarly, that the SRCNN network results in a noticeable boost in the pixelwise accuracy, while the SRGAN performs poorly in comparison. The variance of the dataset PSNR changes when applying SRCNN and SRGAN methods compared to Bicubic methods. For the sandstone dataset, the the highly filtered and smooth Gildehauser sandstone sample results in high SR PSNR values as the smoothness of the image can be inferred by the network, while Bicubic methods cannot. Similarly, the significant reduction in PSNR variance in the Coal samples when using SR methods can be attributed to the presence of thin fracture features that bicubic methods struggle to regenerate, resulting in low PSNR values. 

\begin{table}[htp!]
 \caption{PSNR results for rock types in the validation and testing DeepRock-SR dataset}
 \label{tab:valTestPSNRTable}
  \centering
  \begin{tabular}{|c|c|c|c|c|c|c|}
    \hline
    \multicolumn{1}{|c|}{} & \multicolumn{2}{c|}{Sandstone} & \multicolumn{2}{c|}{Carbonate} & \multicolumn{2}{c|}{Coal} \\
    \hline
     & Mean & Var & Mean & Var  & Mean & Var \\
    \hline
    Bicubic & 25.8822 & 0.6802 & 22.9768 & 1.9896 & 39.8738 & 10.3738\\
    \hline
    SRCNN & 28.5986 & 6.1992 & 24.3879 & 1.7475 & 42.6653 & 3.7016\\
    \hline
    SRGAN & 26.2118 & 8.3945 & 21.8533 & 1.3415 & 40.6061   & 4.4339\\
    \hline
  \end{tabular}
\end{table}

While it is visually clear from Figures \ref{fig:trainValImgs} and \ref{fig:trainValDiffs} that SRGAN images posses improvements in perceptual, high frequency, textural features while maintaining accuracy of the bulk features that are recovered by SRCNN, this cannot be quantified appropriately using averaging metrics such as the PSNR since the generated texture is not a pixelwise match. 

\subsection{External Sample Testing}
\label{sec:External Sample Testing}

Testing of the SRCNN generator on unseen images is expected to result in less error compared to previous SRCNN tests on external samples since the training and validation sets are derived from the larger and more diverse DeepRock-SR dataset. However, the samples used in this section are completely unlike the validation/testing sets, which are unseen subsections of training images. Therefore the images used may exist further away from the manifold of features that the network is trained on. 

\begin{figure}[htp!]
  \centering
    \includegraphics[width=\textwidth]{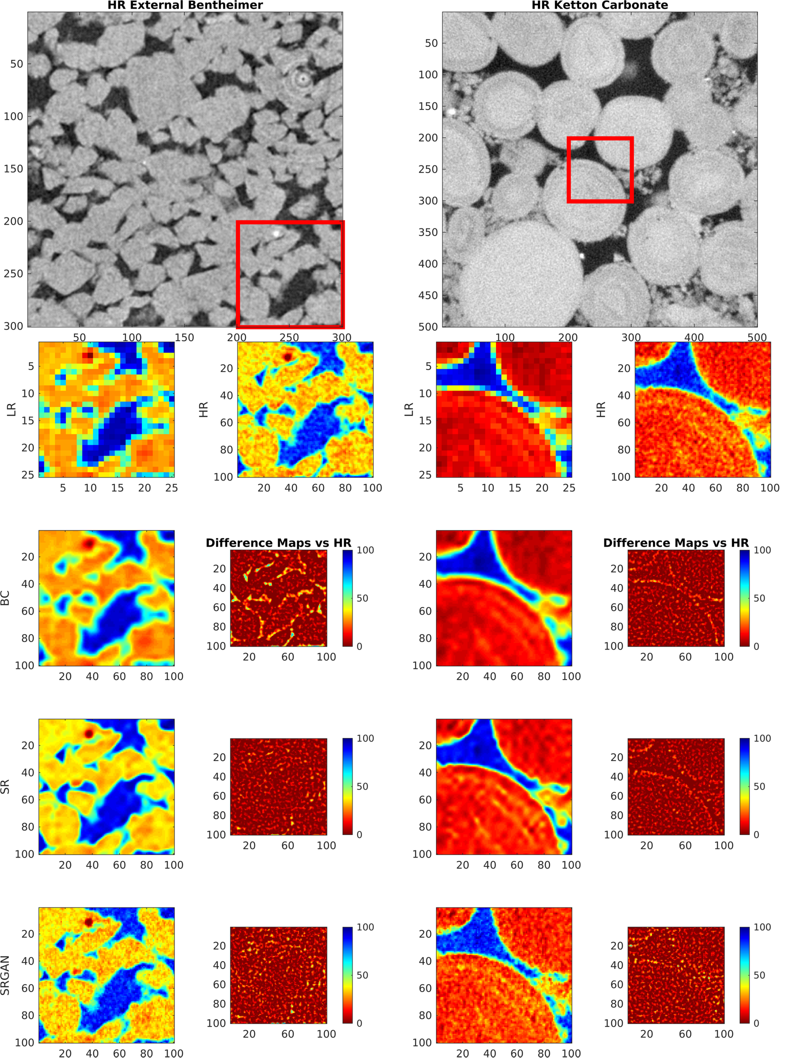}
    \caption{Samples of the external rock images of Bentheimer (left) and Ketton Carbonate (right). Similar characteristics can be observed in terms of the relative performance of bicubic interpolation, SRCNN, and SRGAN. Images show that SRCNN recovers bulk features with high accuracy, while SRGAN regenerates a visually similar textured image to the HR counterpart.}
    \label{fig:extDiffBoxFig}
\end{figure}

\begin{figure}[htp!]
  \centering
    \includegraphics[width=\textwidth]{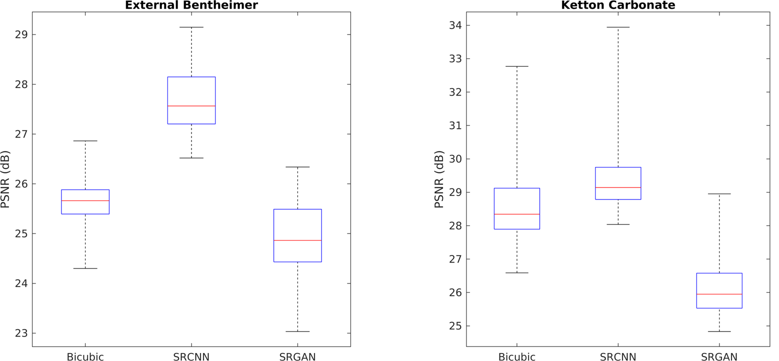}
    \caption{Boxplots of the external rock images of Bentheimer (left) and Ketton Carbonate (right). PSNR pixelwise error shows the typical boost over bicubic interpolation that SRCNN achieves, while SRGAN is less accurate. Overall, these are again in line with expected results obtained during validation and testing in section \ref{sec:Validation and Testing Analysis}.}
    \label{fig:extboxplotFig}
\end{figure}

Testing the network on these completely unseen images shows that SRCNN performance is impacted so a slight degree in terms of the PSNR, seen in Figure \ref{fig:extboxplotFig}, but retains its superiority to BC images in terms of edge feature recovery and overall pixelwise matching, seen in Figure \ref{fig:extDiffBoxFig}. Similarly, SRGAN generated images of the external samples remain visually similar to their HR counterparts. PSNR performance is as expected, with SRCNN improving over bicubic methods, and SRGAN resulting in a decline in pixelwise accuracy. The more globular features of Bentheimer Sandstone and Ketton Carbonate are less sensitive to the edge-loss incurred with bicubic interpolation. 

While the SRCNN and SRGAN generated images for these external samples are significant improvements on interpolation methods in terms of their features and textures, deviation from the ground truth is unavoidable to an extent, as deep learning is highly reliant on large amounts of representative data. Other further deviations can be expected with images of low contrast (not included in the DeepRock-SR dataset) or otherwise. Testing network performance on completely unseen digital rocks inevitably results in a performance penalty that can be rectified by ever increasingly diverse training data.

\subsection{LR Augmentation for Generalisation of Blur and Noise}
\label{sec:Augmentation}

\begin{figure}[htp!]
  \centering
    \includegraphics[width=\textwidth]{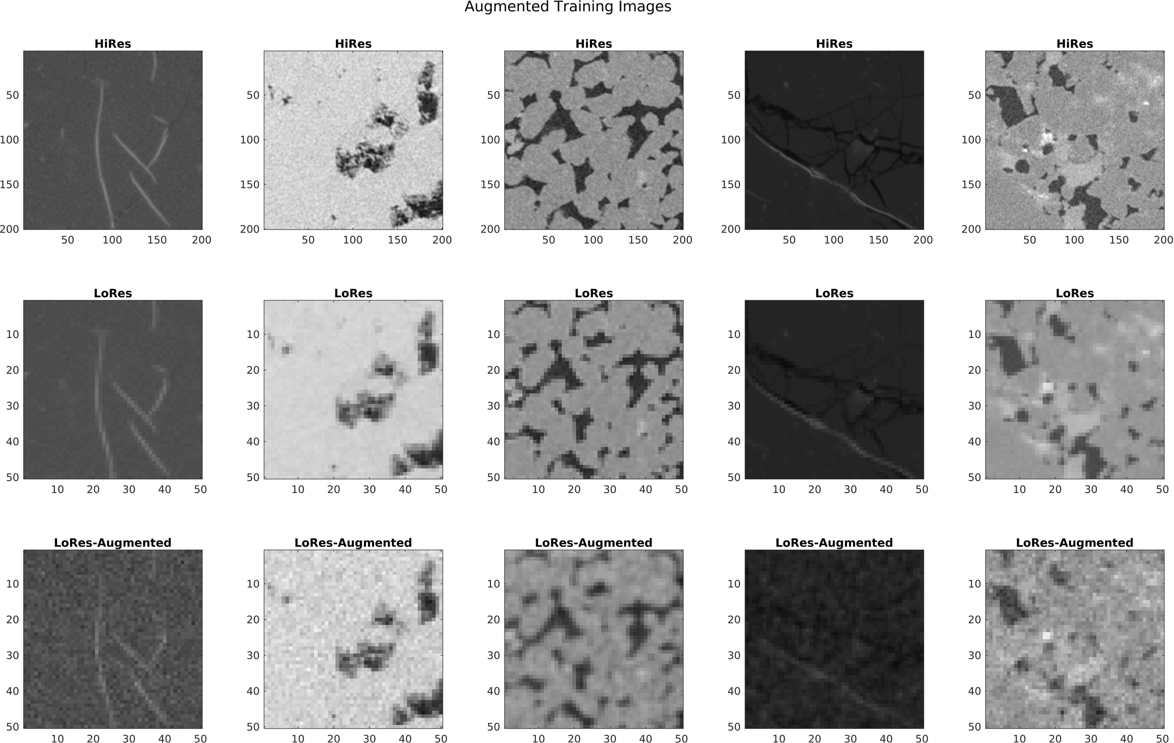}
    \caption{Sample images of the resulting augmented LR image compared to the synthetic LR image, and the original HR image.}
    \label{fig:noisyTrainingComparisonFig}
\end{figure}

\begin{figure}[htp!]
  \centering
    \includegraphics[width=\textwidth]{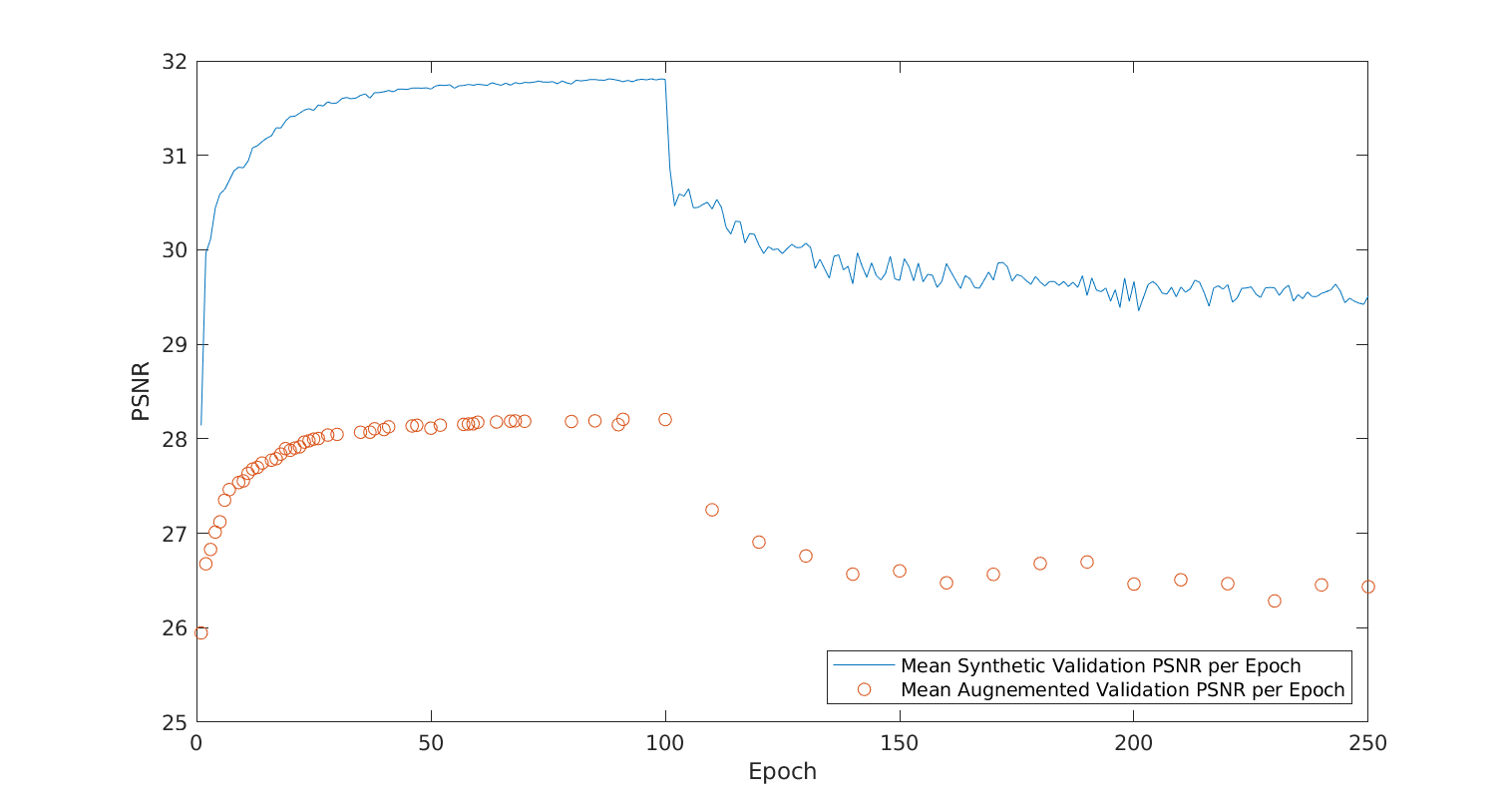}
    \caption{Validation PSNR achieved over training epochs for the augmented dataset training, compared to the synthetic data from Figure \ref{fig:trainValPSNR}. There is a significant drop in network performance, as the augmented LR images provide generalisation to the SR results by adding unlearnable randomness to the LR-HR pairing}
    \label{fig:SynAugTrainValPSNR}
\end{figure}

Training SRCNN and SRGAN networks using synthetic data generated using downsampling kernels ultimately limits the capabilities of the resulting generator when dealing with images that contain characteristics that are not present in synthetic images. In particular, the image noise may be different and there may be some blur associated with the LR image. While maintaining synthetic sample training, it has been shown that augmenting the LR training images with noise and blur will significantly impact SRCNN performance \cite{wang2019super}. Blur is applied as a Gaussian smoothing kernel with a standard deviation randomly sampled from 0 to 1, and noise is applied as Gaussian white noise with mean of 0 and variance randomly sampled between 0 and 0.005. The resulting LR training images are depicted in Figure \ref{fig:noisyTrainingComparisonFig}. Previously published results indicate that application of these augmentation parameters results in SR images that are highly denoised with very sharp and accurate edges. This is due to the increase in the "unlearnable" randomness of the LR to HR mapping. This extreme loss of intraphase detail is beneficial for segmentation of well resolved features such as sandstone grains, but is detrimental to characterising the microporosity and under-resolution features present in coal and carbonate images. Training on these augmented images for the full schedule of 250 epochs of 1000 iterations, with GAN training starting from epoch 101 results in validation PSNR values that are given in Figure \ref{fig:SynAugTrainValPSNR}, showing a clear reduction in SR performance from a pixelwise perspective. The SR images generated from the network trained on synthetic images, and the ones generated from the network trained on augmented images, will hereinafter be referred to simply as "synthetic" and "augmented" SR images

\begin{figure}[htp!]
  \centering
    \includegraphics[width=\textwidth]{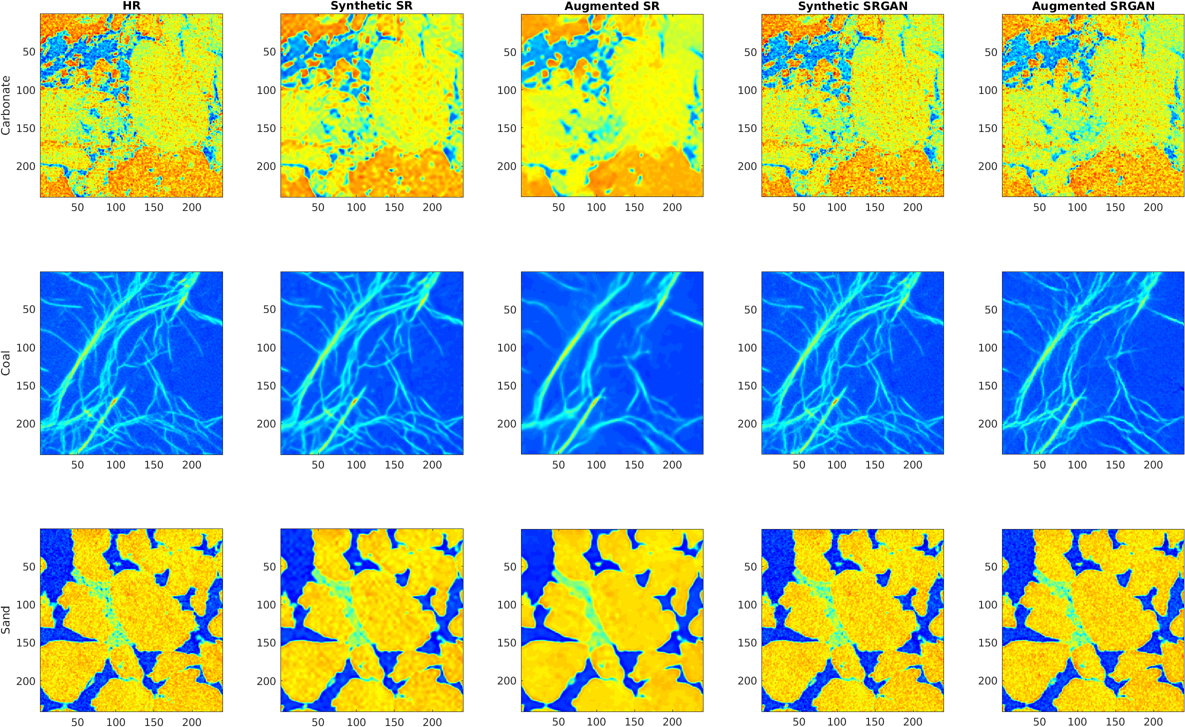}
    \caption{Sample images generated from training with synthetic and augmented datasets. Compared to the synthetic SR images (also shown in Figure \ref{fig:trainValImgs}), augmented SR images provide generalisation to the SR results by adding unlearnable randomness to the LR-HR pairing. SRGAN images on the other hand, remain perceptually convincing regenerations of the HR image, as any losses in the pixelwise recovery of intraphase detail is compensated for by the GAN. The lower contrast coal images are significantly affected by the augmentation, which is likely too excessive for the image characteristics.}
    \label{fig:synAugBoxFig}
\end{figure}

\begin{figure}[htp!]
  \centering
    \includegraphics[width=\textwidth]{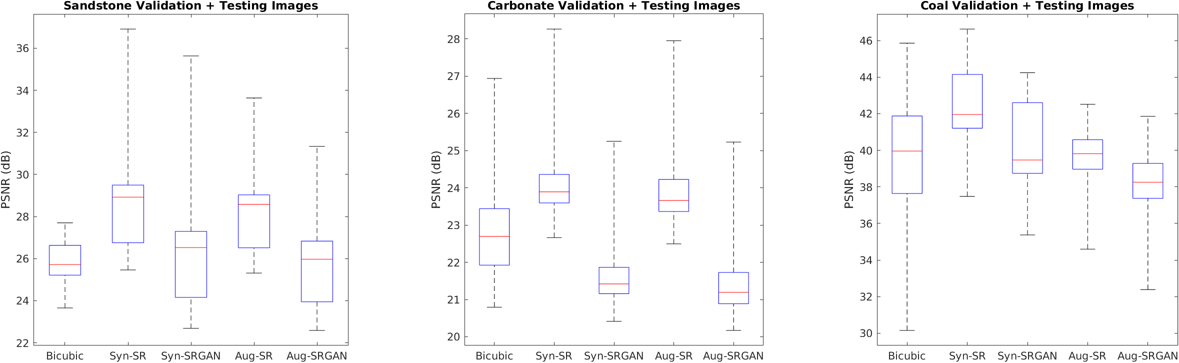}
    \caption{Box plots of the PSNR achieved using synthetic and augmented training datasets. Augmented training results in a reduction in PSNR performance, and in the case of SRGAN images and coal images, lower performance than even bicubic methods. However, a visual inspection of the sample images in Figure \ref{fig:synAugBoxFig} reveals that the losses in the pixelwise accuracy do not occur at the edges of important edge features, but instead occurs in the intraphase regions of the image.}
    \label{fig:AugboxplotFig}
\end{figure}

Example SR images obtained from the synthetic network compared to the augmented network shows in Figure \ref{fig:synAugBoxFig} shows that synthetic images outperform augmented images in feature recovery, though both results retain larger scale edgewise features \cite{wang2019super}. The GAN trained results again show little perceptual difference. There are losses in the detail in coal SR images during augmentation, likely due to an excessive augmentation of noise and blur, that affects the lower contrast coal images more significantly. Boxplots of the PSNR over the 800 validation and testing images of sandstone, carbonate, and coal in Figure \ref{fig:AugboxplotFig} show that the PSNR is affected adversely, with the loss of intraphase features resulting in lower PSNR values than even bicubic methods. The overall PSNR is lower in augmented images, but the final result is compensated by the discriminator to regenerate the texture in a perceptually convincing manner. 

Despite the reduced PSNR achieved by the augmented training, the seemingly high fidelity texture regeneration shows that sandstone and carbonate Aug-SRGAN images result in a similar visual texture match with the Syn-SRGAN images, indicating that the GAN texture that is regenerated is spatially similar to the HR images. 

\begin{figure}[htp!]
  \centering
    \includegraphics[width=\textwidth]{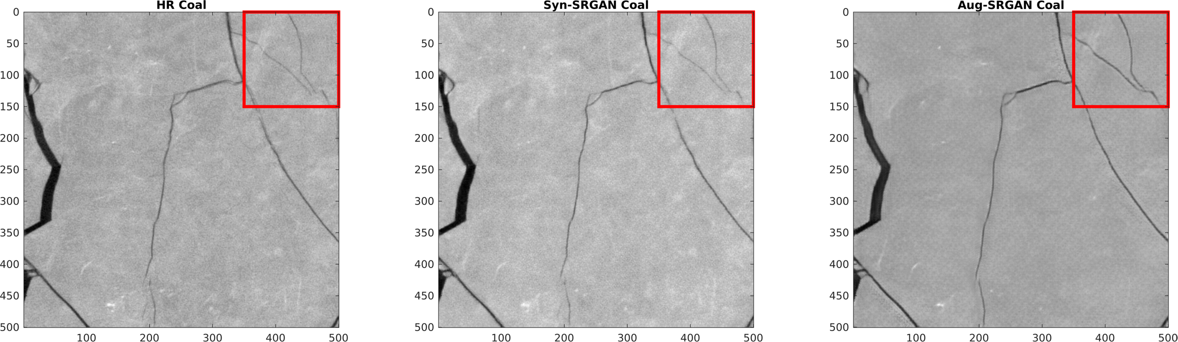}
    \caption{Sample image of how the blur and noise augmentation causes an overestimation of the fracture aperture in coal images, as the fracture features are too attenuated by the augmentation process.}
    \label{fig:coalAugFracFig}
\end{figure}

The lower performance of coal images is due to the presence of thin fracture features that are lost or poorly estimated when the LR images are augmented with blur and noise, seen in a sample image in Figure \ref{fig:coalAugFracFig}, and present throughout the generated augmented SR images. With a less aggressive augmentation, it can be expected that the network will appropriately recover the features and textures. A full repository of the results can be found in conjunction with the DeepRock-SR dataset \cite{DRSRD3}.

\begin{figure}[htp!]
  \centering
    \includegraphics[width=\textwidth]{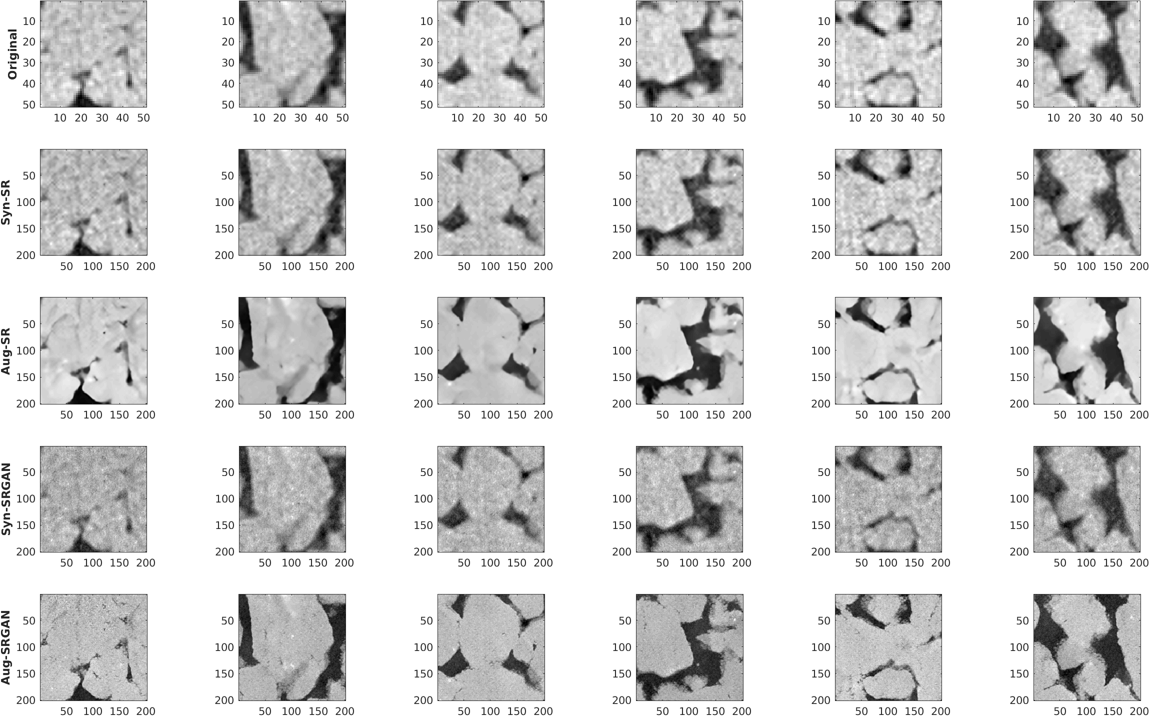}
    \caption{Sample images of the 7 $\mu$m, noisy and blurry Bentheimer sample, with generated SR images with a scaling factor of 4x (resulting in a 1.75 $\mu$m resolution), trained with synthetic and augmented images. It can be seen that the synthetic images do not account for the blur present in the original images, and generate perceptually unsatisfying results. The augmented networks are capable of accounting for image blur and thus generate higher fidelity results.}
    \label{fig:bentRamFig}
\end{figure}

To illustrate and quantify the limitations and different results that can be expected between synthetic and augmented training, we use a Bentheimer sample with a resolution of 7 $\mu$m as a LR input and generate SR images. This sample contains some natural blur and noise that is outside the manifold that synthetic LR images exist on, thus outside the parameters learned from training on synthetically downsampled images. From the original 7 $\mu$m, noisy and blurry image, a few SR images can be generated from the trained networks, shown in Figure \ref{fig:bentRamFig}. The synthetically trained networks generate images that contain higher resolution features and textures, but remain blurry and noisy. The augmented networks instead are capable of sharpening the input image as LR image blur is a recognised input feature and are generally more perceptually superior.

\subsection{Super Resolution of High Resolution Images}
\label{sec:HRSR}

\begin{figure}[htp!]
  \centering
    \includegraphics[width=\textwidth]{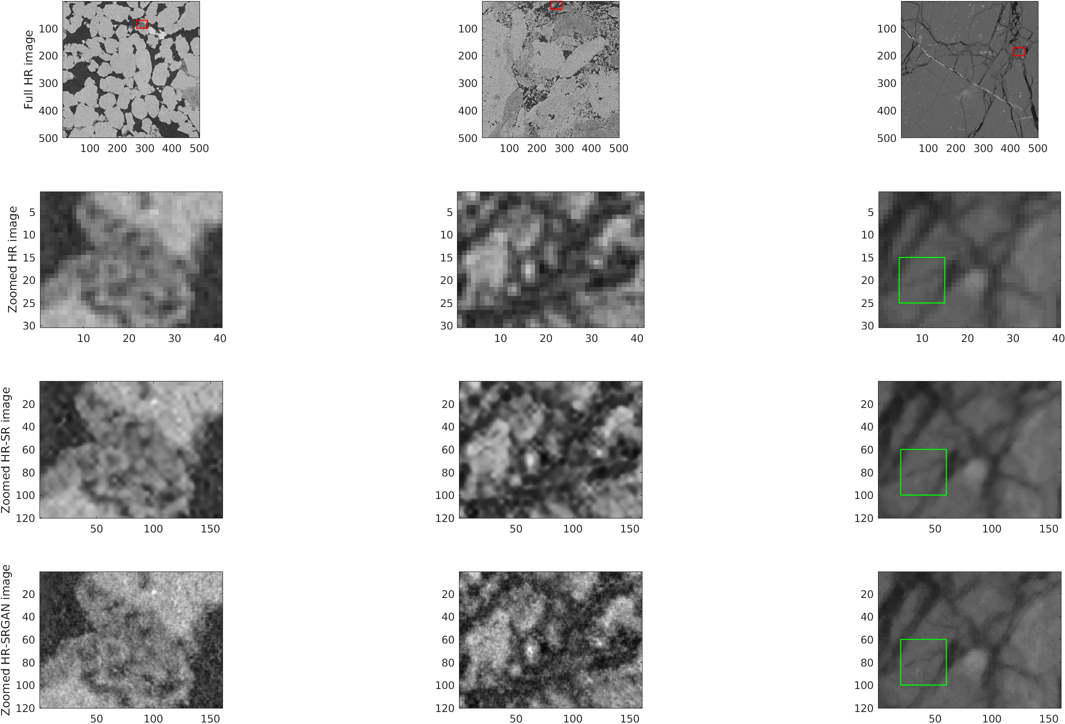}
    \caption{Sample images of HR images in the DeepRock-SR dataset, used as inputs for generating HR-SR images. Features that are near the limit of the image resolution become resolved and textured.}
    \label{fig:HRSRFig}
\end{figure}

Since the network is trained on downsampled LR counterparts (synthetic and augmented) of HR ground truth images, the HR images themselves are unseen by the trained network. Feeding the HR images as input LR images will generate HR-SR images with an image resolution higher than the range that the training is conducted with. The Bentheimer sample 1 \cite{DRSRD1}, Estaillades Carbonate \cite{estCarb}, and Sheared Coal \cite{shearedCoal} samples are used to represent the different rock types of interest, shown in Table \ref{tab:HRSRtable}. The DeepRock-SR dataset contains sandstone and carbonate images with a HR to LR feature mapping of around 3-5 $\mu$m to 12-20 $\mu$m, and a coal mapping of 25 $\mu$m to 100 $\mu$m. Feeding the trained network an input sandstone image of for example, 4 $\mu$m, and expecting an accurate SRGAN image with a resolution of $\mu$m is an extrapolation of performance. The SRCNN and SRGAN synthetic generators are applied to the HR images of each sample, and samples of the HR-SR images are outlined in Table \ref{tab:HRSRtable} and shown in Figure \ref{fig:HRSRFig}. 

\begin{table}[htp!]
 \caption{HR images used as input to generate HR-SR images}
 \label{tab:HRSRtable}
  \centering
  \begin{tabular}{lllll}
    \toprule
    \multicolumn{2}{c}{Part}                   \\
    \cmidrule(r){1-2} 
    Name & HR Resolution ($\mu$m) & HR-SR Resolution ($\mu$m) \\
    \midrule
    Bentheimer Sandstone & 3.8 & 0.95\\
    \midrule
    Estaillades Carbonate & 3.1 & 0.7525 \\
    \midrule
    Sheared Coal & 16 & 4 \\
    \bottomrule
  \end{tabular}
\end{table}

In this case, there are no ground truth images to compare quantitatively with so PSNR metrics of pixelwise accuracy are unavailable to analyse. Of particular interest is the generation of sub-pixel features of micro-porous or poorly resolved regions, such as the partially dissolved mineralisation found in the Bentheimer image, the micro-porous texture of the Estaillades Carbonate, and any under-resolved fractures present in the coal images. These features typically require SEM imaging or nano-CT to resolve properly and will form part of our future work in this area of research. In the generated SR and SRGAN images, the under-resolved features that were present in the original HR images become resolved and sharpened, with texture regenerated by the SRGAN. In the cases of sandstone and carbonate features, SRGAN results are visually sharp, with textures appearing qualitatively as expected in a typical greyscale $\mu$CT image. The coal image on the other hand appears to remain perceptually blurry and undertextured. However, under-resolved fracture features are clearly resolved in the HR-SR image, shown in Figure \ref{fig:HRSRFig} by the green bounding boxes. It is  important to emphasise that by applying the trained network on images with a voxel size mapping (1-5 $\mu$m) that is outside the range of training  (5-20 $\mu$m), the primary assumption is that the type of texture exists at the smaller length scales of the domain. While an exercise in extrapolation, this suggests that if the network (or similar SRGAN architecture) is trained on a wider range of image resolutions (obtained from nano-CT, SEM, or otherwise), performance would remain accurate.

\subsection{Direct Visual Comparison with SEM Images}
\label{sec:HRSRSEM}

\begin{figure}[!htp]
\caption{$\mu$CT image (left) and registered SEM image (right) of Mt Simon slice}
 \label{fig:SEMandLRCT}
  \centering
  \begin{minipage}[b]{0.4\textwidth}
    \includegraphics[width=\textwidth]{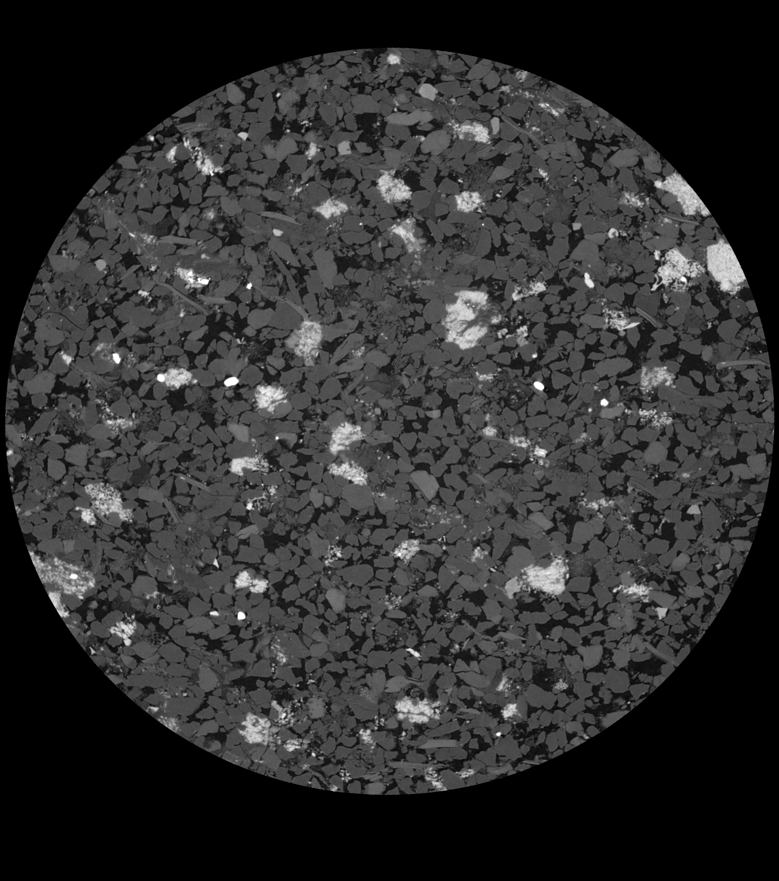}
  \end{minipage}
  \hfill
  \begin{minipage}[b]{0.4\textwidth}
    \includegraphics[width=\textwidth]{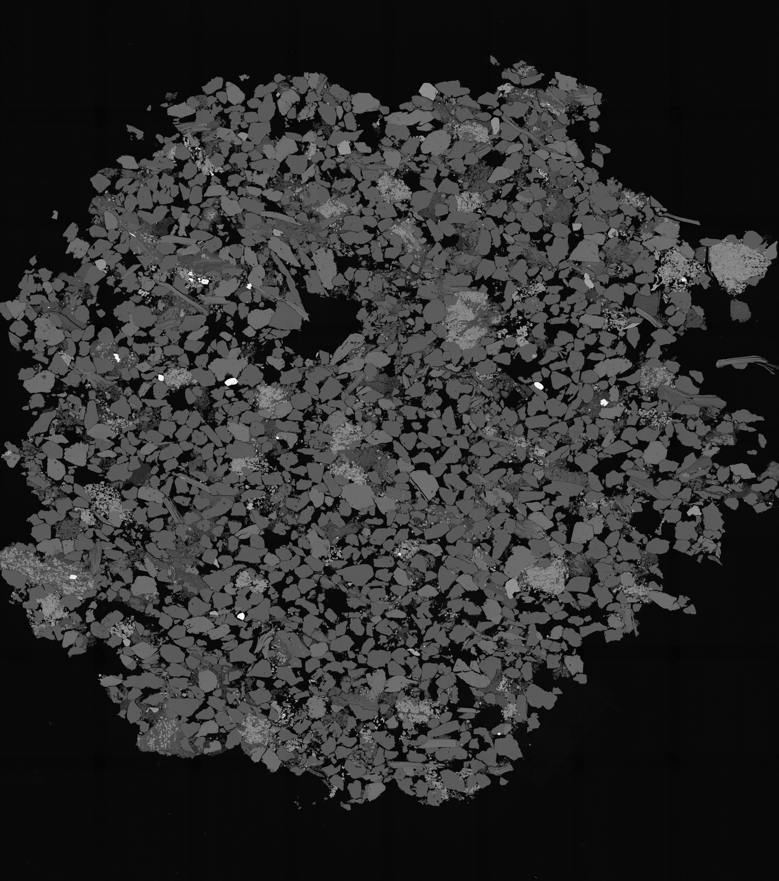}
  \end{minipage}
\end{figure}

A final test is performed using the trained network by taking a SEM scan of a slice from a Mt Simon sandstone and comparing its detail and features with the equivalent $\mu$CT image. The SEM image measures 7047x6226, at a resolution of 0.5 $\mu$m, while the equivalent CT image was imaged at 1.7 $\mu$m and resampled to the suitable 4x image size of 1762x1557, resulting in a voxel size of 2 $\mu$m. These testing images can be seen in Figure \ref{fig:SEMandLRCT}. The lower resolution $\mu$CT image is super resolved to boost the resolution and recover the texture of the image. Once again, this particular use case involves images with voxel sizes outside the range of the DeepRock-SR training dataset. The SEM image characteristics are also very different to the $\mu$CT, as it possesses high image sharpness, showing each pixel discretely unlike the $\mu$CT images which are more diffuse with information spanning multiple pixels. Since the network was trained on such diffuse images, it is not possible under the current training scheme to reach image detail levels in the range of the SEM image.

\begin{figure}[htp!]
  \centering
    \includegraphics[width=\textwidth]{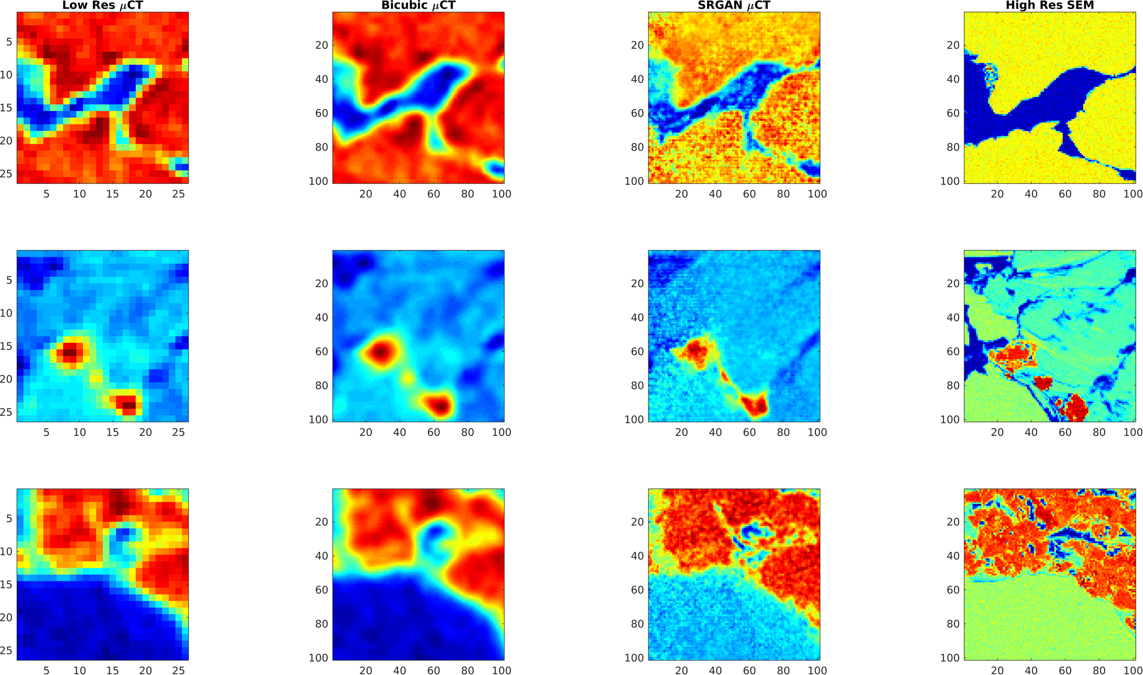}
    \caption{Sample subsections of the Mt Simon sandstone, with comparison between the original low resolution, the bicubic upsample, the SRGAN upsample, and the registered SEM image. There is a clear improvement in image quality achieved by the use of the EDSRGAN network.}
    \label{fig:HRSRSEMFig}
\end{figure}

A selection of small subsets of the images is shown in Figure \ref{fig:HRSRSEMFig}, and shows a clear visual improvement in the regeneration of features from a low resolution source image compared to a bicubic upsampling. There are some convolutional artefacts present in the SRGAN $\mu$CT generated images due to the extrapolative nature of this test case, but the results show that the images generated by EDSRGAN improve connectivity and feature detail in a manner that is visually consistent with the high fidelity SEM images. Aside from convolutional artefacts from network extrapolation, the main sources of comparative error come from (a) the inherently more diffuse quality of the $\mu$CT images that the network is trained on, (b) the $\mu$CT image pixels contain diffuse information from the previous and subsequent slice, and (c) registration error and changes in grain locations between imaging.

\pagebreak

\section{Conclusions}
By training the network with the DeepRock-SR dataset, the inherent hardware limitations with FOV and resolution of Digital Rock Images can be compensated. A low resolution, noisy, blurry image of sandstone, carbonate or coal can be transformed into a high fidelity, accurately textured SR image. This is possible by combining SRCNN that difference maps have shown to excel in capturing edge details with a GAN to recover high frequency texture details. The SRCNN network shows a 3-5 dB boost in pixel accuracy (50\% to 70\% reduction in relative error) compared to bicubic interpolation, while texture shows superior similarity from SRGAN images compared to normal SRCNN and other interpolation methods. Extrapolation of training with external, morphologically distinct samples remains similarly accurate based on pixel error and visual texture analysis. Generalising the LR to HR mapping with augmentation results in a high adaptability to noise and blur. HR-SR images generated by feeding HR images into the network to extrapolate performance to sub-resolution features in the HR images themselves show that under-resolution features are recovered in high detail. Dissolved minerals and thin fractures are regenerated despite the network operating outside of trained specifications. Direct comparison to SEM images of a Mt Simon sandstone show that even in the extrapolative tests, the generated details are consistent with the underlying geometry of the sample, which bicubic interpolations are unable to achieve. 

The neural network architecture used in this study can be made more efficient with wide activation network trimming \cite{WDSR}, or more flexible with unpaired, unsupervised cyclical networks \cite{circleGAN}. The choice of loss function weights will affect the interplay between generator and discriminator (features and textures), and the availability of training data and augmentation further adds avenues of algorithmic improvement. In its current state, the network is operational and adequately effective in its accuracy and flexibility for the purposes of illustrating and quantifying the general effectiveness of SRGAN methods. SRGAN images are accurate in both a feature and texture basis when applied to unseen validation and testing LR images within the boundaries of the dataset used for training, as seen in Figure \ref{fig:trainValDiffs}. However, when there does not exist a ground truth image to calculate comparative metrics with, there is understandable scepticism and suspicion towards the generation of texture within the intra-phase with SRGAN methods, especially true in the case of extrapolative uses of trained networks. The use of 2D super resolution models on 3D data ignores the depth dimension, which limits this study as a proof of concept for the recovery and generation of micro-CT images that can be segmented for digital rock workflows. Application of specialised networks to 3D super resolution of micro-CT images and coupled Super Resolution with Segmentation is natural progression from this work. 

The recovery of both features and texture from LR images is beneficial for characterising digital rocks that are dominated by under-resolution micro-porous features such as in carbonate and coal samples. Overall, images can be constrained by the brittle mineralogy of the rock (coal), by lower quality fast transient imaging (waterflooding), or by the energy of the source (microporosity). These limitations can be overcome and super resolved accurately for further analysis downstream. The neural network architecture and training methodology used in this study provide the tools necessary to generate HR $\mu$CT images that exceed typical imaging limits. 

\section{Acknowledgements}
We would like to acknowledge both the Tyree X-ray facility (\url{http://www.tyreexrayct.unsw.edu.au/}), and the Digital Rocks Portal (\url{https://www.digitalrocksportal.org/projects/}) for providing images that were analysed in this paper. Images used and generated in this paper can be found in the DeepRock-SR repository on the Digital Rocks Portal.


\begin{thebibliography}{10}

\bibitem{lindquist}
W~Brent Lindquist, Sang‐Moon Lee, David~A Coker, Keith~W Jones, and Per
  Spanne.
\newblock Medial axis analysis of void structure in three‐dimensional
  tomographic images of porous media.
\newblock {\em Journal of Geophysical Research: Solid Earth},
  101(B4):8297--8310, 1996.

\bibitem{Hazlett}
RD~Hazlett.
\newblock {\em Simulation of capillary-dominated displacements in
  microtomographic images of reservoir rocks}, pages 21--35.
\newblock Springer, 1995.

\bibitem{Wildenschild}
Dorthe Wildenschild and Adrian~P Sheppard.
\newblock X-ray imaging and analysis techniques for quantifying pore-scale
  structure and processes in subsurface porous medium systems.
\newblock {\em Advances in Water Resources}, 51:217--246, 2013.

\bibitem{RN7}
Steffen Schlüter, Adrian Sheppard, Kendra Brown, and Dorthe Wildenschild.
\newblock Image processing of multiphase images obtained via x-ray
  microtomography: A review.
\newblock {\em Water Resources Research}, 50(4):3615--3639, 2014.

\bibitem{DGDD}
Ying~Da Wang, Traiwit Chung, Ryan~T. Armstrong, James~E. McClure, and Peyman
  Mostaghimi.
\newblock Computations of permeability of large rock images by dual grid domain
  decomposition.
\newblock {\em Advances in Water Resources}, 126:1--14, 2019.

\bibitem{pfvs}
T.~Chung, Y.D. Wang, Mostaghimi P., and Armstrong RT.
\newblock Approximating permeability of micro-ct images using elliptic flow
  equations.
\newblock {\em SPE Journal}, 2018.

\bibitem{wang2019multi}
Ying~Da Wang, Traiwit Chung, Ryan~T. Armstrong, James McClure, Thomas Ramstad,
  and Peyman Mostaghimi.
\newblock Accelerated computation of relative permeability by coupled
  morphological-direct multiphase flow simulation, 2019.

\bibitem{peymanK}
Peyman Mostaghimi, Martin~J Blunt, and Branko Bijeljic.
\newblock Computations of absolute permeability on micro-ct images.
\newblock {\em Mathematical Geosciences}, 45(1):103--125, 2013.

\bibitem{Krakowska}
Paulina Krakowska, Marek Dohnalik, Jadwiga Jarzyna, and Kamila Wawrzyniak-Guz.
\newblock Computed x-ray microtomography as the useful tool in petrophysics: A
  case study of tight carbonates modryn formation from poland.
\newblock {\em Journal of Natural Gas Science and Engineering}, 31:67--75,
  2016.

\bibitem{ferrand1992effect}
Lin~A Ferrand and Michael~A Celia.
\newblock The effect of heterogeneity on the drainage capillary
  pressure-saturation relation.
\newblock {\em Water Resources Research}, 28(3):859--870, 1992.

\bibitem{Flannery}
Brian~P Flannery, Harry~W Deckman, Wayne~G Roberge, and Kevin~L d'Amico.
\newblock Three-dimensional x-ray microtomography.
\newblock {\em Science}, 237(4821):1439--1444, 1987.

\bibitem{Coenen}
J~Coenen, E~Tchouparova, and X~Jing.
\newblock Measurement parameters and resolution aspects of micro x-ray
  tomography for advanced core analysis.
\newblock In {\em proceedings of International Symposium of the Society of Core
  Analysts}.

\bibitem{jiang2013representation}
Zeyun Jiang, MIJ Van~Dijke, Kenneth~Stuart Sorbie, and Gary~Douglas Couples.
\newblock Representation of multiscale heterogeneity via multiscale pore
  networks.
\newblock {\em Water resources research}, 49(9):5437--5449, 2013.

\bibitem{schluter2014image}
Steffen Schl{\"u}ter, Adrian Sheppard, Kendra Brown, and Dorthe Wildenschild.
\newblock Image processing of multiphase images obtained via x-ray
  microtomography: a review.
\newblock {\em Water Resources Research}, 50(4):3615--3639, 2014.

\bibitem{yi2017pore}
Zhixing Yi, Mian Lin, Wenbin Jiang, Zhaobin Zhang, Haishan Li, and Jian Gao.
\newblock Pore network extraction from pore space images of various porous
  media systems.
\newblock {\em Water Resources Research}, 53(4):3424--3445, 2017.

\bibitem{Li_and_Teng}
Zhengji Li, Qizhi Teng, Xiaohai He, Guihua Yue, and Zhengyong Wang.
\newblock Sparse representation-based volumetric super-resolution algorithm for
  3d ct images of reservoir rocks.
\newblock {\em Journal of Applied Geophysics}, 144:69--77, 2017.

\bibitem{BULTREYS201536}
Tom Bultreys, Luc~Van Hoorebeke, and Veerle Cnudde.
\newblock Multi-scale, micro-computed tomography-based pore network models to
  simulate drainage in heterogeneous rocks.
\newblock {\em Advances in Water Resources}, 78:36 -- 49, 2015.

\bibitem{coalFragility}
Wen Hao~Chen, Y~Yang, Tiqiao Xiao, Sheridan Mayo, Yu~Dan~Wang, and Haipeng
  Wang.
\newblock A synchrotron-based local computed tomography combined with
  data-constrained modelling approach for quantitative analysis of anthracite
  coal microstructure.
\newblock {\em Journal of synchrotron radiation}, 21:586--93, 05 2014.

\bibitem{wang2019super}
Ying~Da Wang, Ryan Armstrong, and Peyman Mostaghimi.
\newblock Super resolution convolutional neural network models for enhancing
  resolution of rock micro-ct images, 2019.

\bibitem{sr2003overview}
S.~C. Park, M.~K. Park, and M.~G. Kang.
\newblock Super-resolution image reconstruction: a technical overview.
\newblock {\em IEEE Signal Processing Magazine}, 20(3):21--36, May 2003.

\bibitem{SRCNNDong}
Chao Dong, Chen~Change Loy, Kaiming He, and Xiaoou Tang.
\newblock {\em Image Super-Resolution Using Deep Convolutional Networks},
  volume~38.
\newblock 2014.

\bibitem{softCuts}
S.~{Dai}, M.~{Han}, W.~{Xu}, Y.~{Wu}, Y.~{Gong}, and A.~K. {Katsaggelos}.
\newblock Softcuts: A soft edge smoothness prior for color image
  super-resolution.
\newblock {\em IEEE Transactions on Image Processing}, 18(5):969--981, May
  2009.

\bibitem{gradProfPrior}
{Jian Sun}, {Zongben Xu}, and {Heung-Yeung Shum}.
\newblock Image super-resolution using gradient profile prior.
\newblock In {\em 2008 IEEE Conference on Computer Vision and Pattern
  Recognition}, pages 1--8, June 2008.

\bibitem{gradProfSharp}
Q.~{Yan}, Y.~{Xu}, X.~{Yang}, and T.~Q. {Nguyen}.
\newblock Single image superresolution based on gradient profile sharpness.
\newblock {\em IEEE Transactions on Image Processing}, 24(10):3187--3202, Oct
  2015.

\bibitem{MRFExmaple}
W.~T. {Freeman}, T.~R. {Jones}, and E.~C. {Pasztor}.
\newblock Example-based super-resolution.
\newblock {\em IEEE Computer Graphics and Applications}, 22(2):56--65, March
  2002.

\bibitem{neighbourEmbedding}
{Hong Chang}, {Dit-Yan Yeung}, and {Yimin Xiong}.
\newblock Super-resolution through neighbor embedding.
\newblock In {\em Proceedings of the 2004 IEEE Computer Society Conference on
  Computer Vision and Pattern Recognition, 2004. CVPR 2004.}, volume~1, pages
  I--I, June 2004.

\bibitem{sparseRep}
Jianchao Yang, John Wright, Thomas S.~Huang, and Lei Yu.
\newblock Image super-resolution via sparse representation.
\newblock {\em Image Processing, IEEE Transactions on}, 19:2861 -- 2873, 12
  2010.

\bibitem{randomForest}
S.~{Schulter}, C.~{Leistner}, and H.~{Bischof}.
\newblock Fast and accurate image upscaling with super-resolution forests.
\newblock In {\em 2015 IEEE Conference on Computer Vision and Pattern
  Recognition (CVPR)}, pages 3791--3799, June 2015.

\bibitem{unifiedSR}
J.~{Yu}, X.~{Gao}, D.~{Tao}, X.~{Li}, and K.~{Zhang}.
\newblock A unified learning framework for single image super-resolution.
\newblock {\em IEEE Transactions on Neural Networks and Learning Systems},
  25(4):780--792, April 2014.

\bibitem{WangEEDS}
Yifan Wang, Lijun Wang, Hongyu Wang, and Peihua Li.
\newblock End-to-end image super-resolution via deep and shallow convolutional
  networks.
\newblock {\em CoRR}, abs/1607.07680, 2016.

\bibitem{EDSR}
Bee Lim, Sanghyun Son, Heewon Kim, Seungjun Nah, and Kyoung Mu~Lee.
\newblock {\em Enhanced Deep Residual Networks for Single Image
  Super-Resolution}.
\newblock 2017.

\bibitem{WDSR}
Jiahui Yu, Yuchen Fan, Jianchao Yang, Ning Xu, Zhaowen Wang, Xinchao Wang, and
  Thomas Huang.
\newblock Wide activation for efficient and accurate image super-resolution.
\newblock {\em CoRR}, abs/1808.08718, 2018.

\bibitem{VDSR}
Jiwon Kim, Jung Kwon~Lee, and Kyoung Mu~Lee.
\newblock {\em Accurate Image Super-Resolution Using Very Deep Convolutional
  Networks}.
\newblock 2015.

\bibitem{SRGANledig}
Christian Ledig, Lucas Theis, Ferenc Huszar, Jose Caballero, Andrew Cunningham,
  Alejandro Acosta, Andrew Aitken, Alykhan Tejani, Johannes Totz, Zehan Wang,
  and Wenzhe Shi.
\newblock {\em Photo-Realistic Single Image Super-Resolution Using a Generative
  Adversarial Network}.
\newblock 2017.

\bibitem{umeharaSRCNNmedical}
Kensuke Umehara, Junko Ota, and Takayuki Ishida.
\newblock {\em Application of Super-Resolution Convolutional Neural Network for
  Enhancing Image Resolution in Chest CT}, volume~31.
\newblock 2018.

\bibitem{circleGAN}
Chenyu {You}, Guang {Li}, Yi~{Zhang}, Xiaoliu {Zhang}, Hongming {Shan},
  Shenghong {Ju}, Zhen {Zhao}, Zhuiyang {Zhang}, Wenxiang {Cong}, Michael~W.
  {Vannier}, Punam~K. {Saha}, and Ge~{Wang}.
\newblock {CT Super-resolution GAN Constrained by the Identical, Residual, and
  Cycle Learning Ensemble(GAN-CIRCLE)}.
\newblock {\em arXiv e-prints}, page arXiv:1808.04256, Aug 2018.

\bibitem{wang3DSRCNN}
Yukai Wang, Qizhi Teng, Xiaohai He, Junxi Feng, and Tingrong Zhang.
\newblock Ct-image super resolution using 3d convolutional neural network.
\newblock {\em CoRR}, abs/1806.09074, 2018.

\bibitem{FSRCNN}
Chao Dong, Chen~Change Loy, and Xiaoou Tang.
\newblock Accelerating the super-resolution convolutional neural network.
\newblock {\em CoRR}, abs/1608.00367, 2016.

\bibitem{subpixelConv}
Wenzhe Shi, Jose Caballero, Ferenc Husz{\'{a}}r, Johannes Totz, Andrew~P.
  Aitken, Rob Bishop, Daniel Rueckert, and Zehan Wang.
\newblock Real-time single image and video super-resolution using an efficient
  sub-pixel convolutional neural network.
\newblock {\em CoRR}, abs/1609.05158, 2016.

\bibitem{deconvCheckerboard}
Augustus Odena, Vincent Dumoulin, and Chris Olah.
\newblock Deconvolution and checkerboard artifacts.
\newblock {\em Distill}, 2016.

\bibitem{cycleGAN}
Jun{-}Yan Zhu, Taesung Park, Phillip Isola, and Alexei~A. Efros.
\newblock Unpaired image-to-image translation using cycle-consistent
  adversarial networks.
\newblock {\em CoRR}, abs/1703.10593, 2017.

\bibitem{perceptLoss}
Alexey Dosovitskiy and Thomas Brox.
\newblock Generating images with perceptual similarity metrics based on deep
  networks.
\newblock {\em CoRR}, abs/1602.02644, 2016.

\bibitem{perceptLoss2}
Justin Johnson, Alexandre Alahi, and Fei{-}Fei Li.
\newblock Perceptual losses for real-time style transfer and super-resolution.
\newblock {\em CoRR}, abs/1603.08155, 2016.

\bibitem{ganGoodfellow}
Ian~J. {Goodfellow}, Jean {Pouget-Abadie}, Mehdi {Mirza}, Bing {Xu}, David
  {Warde-Farley}, Sherjil {Ozair}, Aaron {Courville}, and Yoshua {Bengio}.
\newblock {Generative Adversarial Networks}.
\newblock {\em arXiv e-prints}, page arXiv:1406.2661, Jun 2014.

\bibitem{iassonov2009segmentation}
Pavel Iassonov, Thomas Gebrenegus, and Markus Tuller.
\newblock Segmentation of x-ray computed tomography images of porous materials:
  A crucial step for characterization and quantitative analysis of pore
  structures.
\newblock {\em Water Resources Research}, 45(9), 2009.

\bibitem{DRSRD3}
Ying~Da Wang, Peyman Mostaghimi, and Ryan Armstrong.
\newblock A diverse super resolution dataset of sandstone, carbonate, and coal
  (deeprock-sr), 2019.

\bibitem{nonLocal}
A.~{Buades}, B.~{Coll}, and J.~. {Morel}.
\newblock A non-local algorithm for image denoising.
\newblock In {\em 2005 IEEE Computer Society Conference on Computer Vision and
  Pattern Recognition (CVPR'05)}, volume~2, pages 60--65 vol. 2, June 2005.

\bibitem{DRSRD1}
Ying~Da Wang, Peyman Mostaghimi, and Ryan Armstrong.
\newblock A super resolution dataset of digital rocks (drsrd1): Sandstone and
  carbonate, 2019.

\bibitem{herringSands}
Anna Herring, Adrian Sheppard, Michael Turner, and Levi Beeching.
\newblock Multiphase flows in sandstones, 2018.

\bibitem{glideSand}
Steffen Berg, Ryan Armstrong, and Andreas Wiegmann.
\newblock Gildehauser sandstone, 2018.

\bibitem{wilcox}
Ayaz Mehmani, Masa Prodanovic, and Kitty~L. Milliken.
\newblock Wilcox tight gas sandstone, 2015.

\bibitem{estCarb}
Tom Bultreys.
\newblock Estaillades carbonate \#2, 2016.

\bibitem{savCarb}
Tom Bultreys.
\newblock Savonnières carbonate, 2016.

\bibitem{massCarb}
Tom Bultreys.
\newblock Massangis jaune carbonate, 2016.

\bibitem{shearedCoal}
D~Nicolas Espinoza.
\newblock Sheared coal sample, 2015.

\bibitem{fracCoal}
D~Nicolas Espinoza.
\newblock Naturally fractured coal sample, 2015.

\bibitem{Guan2019}
Kelly~M. Guan, Marfa Nazarova, Bo~Guo, Hamdi Tchelepi, Anthony~R. Kovscek, and
  Patrice Creux.
\newblock Effects of image resolution on sandstone porosity and permeability as
  obtained from x-ray microscopy.
\newblock {\em Transport in Porous Media}, 127(1):233--245, Mar 2019.

\bibitem{bentRam}
Thomas Ramstad.
\newblock Bentheimer micro-ct with waterflood, 2018.

\bibitem{kettonCarb}
Alessio Scanziani, Kamaljit Singh, and Martin Blunt.
\newblock Water-wet three-phase flow micro-ct tomograms, 2018.

\bibitem{prelu}
Kaiming He, Xiangyu Zhang, Shaoqing Ren, and Jian Sun.
\newblock Delving deep into rectifiers: Surpassing human-level performance on
  imagenet classification.
\newblock {\em CoRR}, abs/1502.01852, 2015.

\bibitem{AdamKingma}
Diederik Kingma and Jimmy Ba.
\newblock {\em Adam: A Method for Stochastic Optimization}.
\newblock 2014.


\end{thebibliography}
\end{document}